\documentclass[11pt,a4paper]{article}
\pdfoutput=1
\usepackage{jinstpub}
\usepackage{setspace}
\usepackage{xcolor}
\usepackage{graphicx}
\usepackage{enumerate}
\usepackage{mathrsfs}
\usepackage{amsthm, amsmath, amssymb,xspace,graphics}
\usepackage{algpseudocode}
\usepackage{algorithm}
\usepackage{amsmath}
\usepackage{tabularx}
\usepackage{booktabs}
\usepackage{capt-of}

\newcommand{\pro}{_{\text{\tiny prop}}}
\newcommand{\obs}{_{\text{\tiny obs}}}

\title{Machine Learning Accelerated Likelihood-Free Event Reconstruction in Dark Matter Direct Detection}

\author[a, 1]{U. Simola,\note{Corresponding author.}}
\author[b, 1]{B. Pelssers,}
\author[b]{D. Barge,}
\author[b]{J. Conrad,}
\author[a]{J. Corander}

\affiliation[a]{University of Helsinki, Department of Mathematics and Statistics, Pietari Kalmin katu 5, Helsinki, Finland}
\affiliation[b]{Stockholm University, Department of Physics, Roslagstullsbacken 21 A, Stockholm, Sweden}

\emailAdd{umberto.simola@helsinki.fi, bart.pelssers@fysik.su.se}

\abstract{Reconstructing the position of an interaction for any dual--phase time projection chamber (TPC) with the best precision is key to directly detecting Dark Matter. Using the likelihood--free framework, a new algorithm to reconstruct the  2--D ($x,y$) position and the size of the charge signal ($e$) of an interaction is presented. The algorithm uses the secondary scintillation light distribution (S2) obtained by simulating events using a waveform generator. To deal with the computational effort required by the likelihood--free approach, we employ the Bayesian Optimization for Likelihood--Free Inference (BOLFI) algorithm. Together with BOLFI, prior distributions for the parameters of interest ($x,y,e$) and highly informative discrepancy measures to perform the analyses are introduced. We evaluate the quality of the proposed algorithm by a comparison against the currently existing alternative methods using a large-scale simulation study. BOLFI provides a natural probabilistic uncertainty measure for the reconstruction and it improved the accuracy of the reconstruction over the next best algorithm by up to $15\%$ when focusing on events at large radii ($R > 30 \text{ cm}$, the outer $37\%$ of the detector). In addition, BOLFI provides the smallest uncertainties among all the tested methods.}

\begin{document}
\maketitle
\flushbottom

\section{Introduction}\label{introduction}

While there is considerable evidence for existence of dark matter coming from astronomical and cosmological observations on different scales, its direct observation remains a challenge \citep{kolb1990early, corbelli2000extended, dodelson2003modern, markevitch2004direct, roszkowski2018wimp}. Several theoretical models have been proposed assuming that dark matter is composed of yet undetected particles \cite{bertone2005g}, with Weakly Interacting Massive Particle (WIMP) being one of the most popular candidates \citep{jungman1996supersymmetric, strigari2013strigari, undagoitia2015dark}. In case of WIMPs, this is in principle possible through observation of recoiling nuclei \citep{goodman1985detectability, undagoitia2015dark}.

As is the case in any rare event search, to achieve a high sensitivity to signal events (nuclear recoil events from WIMPs) the reduction of background events is crucial. Background events in Dark Matter Direct Detection experiments are electronic or nuclear recoils from beta particles, gamma photons or neutrons from a variety of sources. Dual--phase Time Projection Chambers (TPCs), using high density detector media such as liquid xenon (LXe), have been particularly successful in exploring increasing regions of the WIMP parameter space. The TPC design, in combination with the xenon target, provides two main ways to reduce background events. The first way is the ability to infer the type of recoil (electronic or nuclear) by accurately reconstructing the size of the light and charge signal, respectively defined as the number of prompt scintillation photons and ionization electrons. Properly retrieving these quantities allows for the reduction of electronic recoil background events in the analysis. The second way is the ability to reconstruct the 3--D spatial position of events. Most of the background events will interact in the outer region of the TPC due to the self-shielding properties of the high density liquid xenon. By selecting only events from an inner region of the detector, the signal over background ratio is increased. This motivates the need for accurate position and size of the charge signal reconstruction. Position reconstruction is an important part of experiments such as LUX~\citep{Akerib:2017riv} and XENON~\citep{aprile2017xenon1t, pelssers2015position} and relies on the prompt scintillation light, labeled as S1 signal, and the secondary scintillation light generated by the ionization electrons, labeled as S2 signal, both described in Section \ref{dark.matter}.




Defined  $z$ as the depth of the interaction, the goal of position reconstruction is to infer the 3--D (x,y,z) position of the interaction by combining the 2--D (x,y) position of the S2 signal with the depth of the interaction, the latter being estimated by measuring the drift time. The estimation of $z$ is in general easier and more precise than the 2--D (x,y) position reconstruction. With this regard, one important assumption is that the drift field is uniform throughout the TPC. An uniform drift field ensures that electrons drift up straight, so that the 2--D (x,y) position at which they are generated is the same as where they are extracted into the gas phase and produce the S2 signal. When this assumption does not hold, the field distortion needs to be modelled and the reconstructed position needs to be corrected for this distortion. As will be shown in Section~\ref{examples}, the method introduced in the present work directly infers the 2--D (x,y) position without the need for field distortion correction.
Moreover, position reconstruction typically involves reconstructing the  2--D (x,y) position from the S2 hit pattern and then combining that with the drift time in order to find the 3--D (x,y,z) position of the interaction. However, there are also cases where estimating the z coordinate from the S2 signal alone is desirable. In the case of a S2--only driven analysis (where only S2 signals are used, allowing for a lower energy threshold) no drift time is in general available \citep{Aprile:2016wwo}. In fact, the shape of the S2 signals might be used to provide information on the depth of the interaction but is typically an order of magnitude less accurate than using the drift time and is not used for detector fiducialization. With the proposed method the drift time can also in principle be estimated for cases in which only the S2 signal is available.
Finally, in order to reconstruct the recoil energy of an event the number of prompt scintillation photons and the number of ionization electrons need to be determined. In the case of ionization electrons, this involves finding the total charge signal of the S2 event (number of photo--electrons observed), and then applying various corrections, taking into account the position of the interaction. In the last example of Section~\ref{examples} the charge signal of the event is directly inferred from the waveform and from the S2 hit pattern, using as well the bottom PMT array in order to retrieve additional information. 

In the present work we implement a novel position and size of the charge signal reconstruction algorithm, based on the likelihood--free framework. The presented approach is applicable to any dual--phase TPC. To provide concrete examples and simulations we use the specifications of the XENON1T TPC and employ the open source Processor for Analyzing XENON (PAX)~\citep{xenon_collaboration_2018_1195785} data processor and waveform simulator developed by the XENON collaboration.

The article is structured as follows: in Section~\ref{dark.matter} we present the detector physics of direct detection as well as some of the already available algorithms used for reconstructing the position of an event in dual--phase TPCs starting from the S2 hit pattern. In Section~\ref{stat.method} we introduce the likelihood--free framework, justifying the reasons for its recent success, defining the quantities needed in order to perform the analyses and motivating its use in position reconstruction of events in a dual--phase TPC. Several advantages can be obtained by using the likelihood--free framework, such as the availability of a posterior distribution for the parameters of interest as well as retrieving credible intervals straightforwardly from those posteriors. In order to deal with the computationally expensive calculations, the Bayesian Optimization for Likelihood--Free Inference (BOLFI) \citep{gutmann2016bayesian} is introduced and properly designed for executing the proposed analyses. In Section~\ref{examples} we present three examples to explain how the likelihood--free approach works and to show its advantages over the currently existing alternative methods. The first inceptive example uses a simplified model to generate simulated data. For the other two examples, we use the full event simulator PAX processor.
In particular, for the third and final example, beyond reconstructing the  2--D (x,y) position we additionally reconstruct the size of the charge signal ($e$) of the event. Our conclusions are presented in Section~\ref{conclusions}.

\section{XENON1T detector}\label{dark.matter}

Dark Matter direct detection experiments aim to detect Dark Matter from particles scattering on the detector medium. Both nuclear and electronic recoils lead to three physical processes, known respectively as phonon emission, scintillation and ionization, where the proportion for each process is different depending on the type of recoil. By using a detector which is sensitive to two of these processes, nuclear and electronic recoils can be discriminated. Since most WIMP signal models only predict nuclear scattering, the removal of electronic recoil events greatly reduces the number of background events~\cite{bertone_silk_2010}. 

Among the different available types of detectors typically used for direct detection experiments \cite{undagoitia2015dark}, in the present work we focus on noble--liquid TPCs \citep{aprile2011design, aprile2012xenon100, aprile2017xenon1t}, although the method presented in this work can be applied to any liquid--gas TPC, as shown in the following. A dual--phase (liquid--gas) TPC uses both the scintillation and the ionization signals to detect particles scattering on atoms in the detector. Besides providing the ability to distinguish between nuclear and electronic recoil events, a dual--phase TPC can also reconstruct the spatial position of the events. Properly reconstructing the 3--D ($x,y,z$) position of events is crucial in order to discard background events at the edges of the TPC and to perform spatially dependent corrections that are caused by nonuniform detector responses. The ratio between S1 and S2 is used to discriminate between nuclear and electronic recoils, further reducing the background.



\begin{figure}[htbp]
    \begin{center}
        \includegraphics[width=0.8\textwidth]{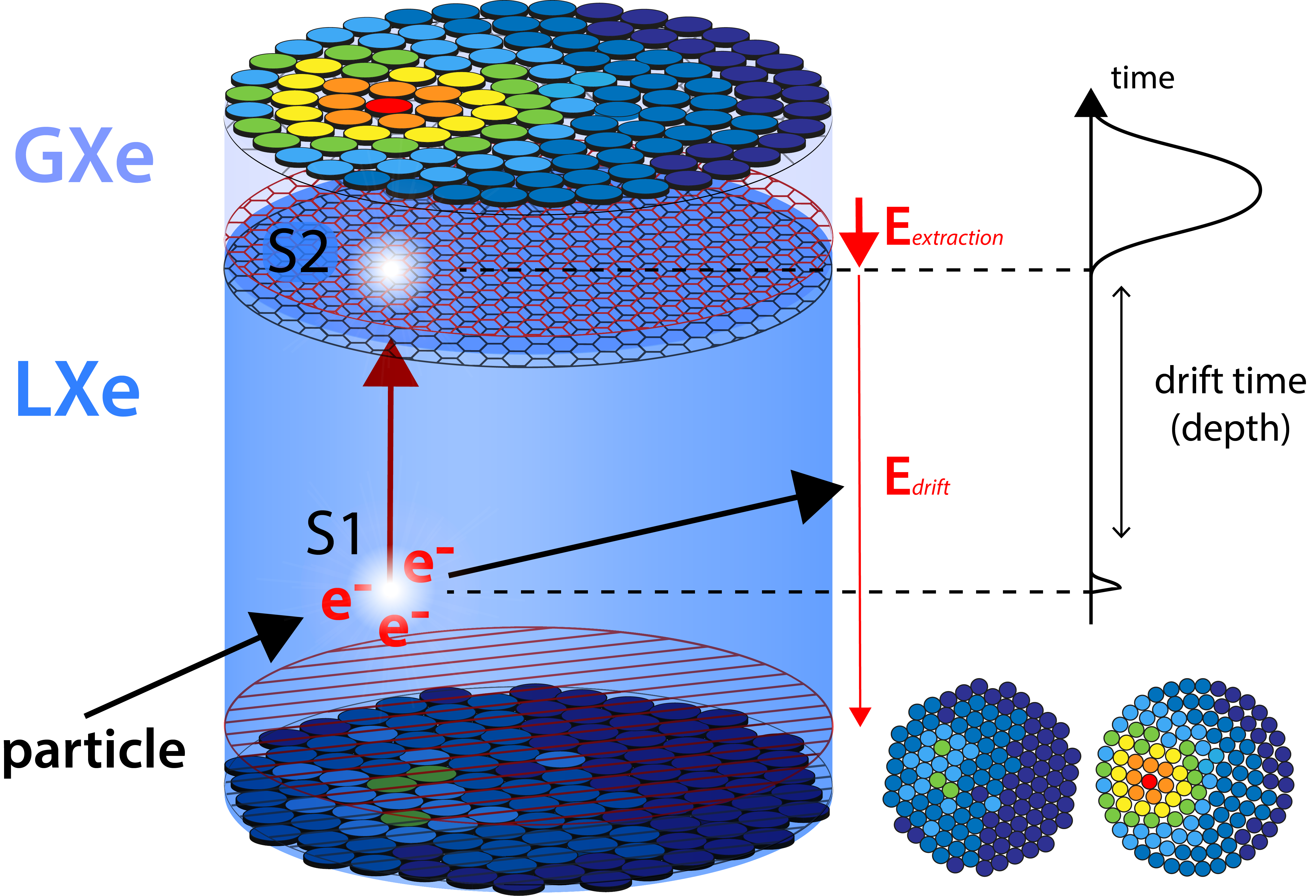}
        \caption{The working principle of a dual--phase (liquid--gas) TPC. A particle interacting with the detector medium will produce scintillation light and ionization charge. The light is seen by the PMTs at the top and bottom of the TPC (the S1 signal). The charges drift to the top of the TPC by the electric field $E_{\text{drift}}$. At the top of the TPC the charges are extracted by the extraction field $E_{\text{extraction}}$ and causing a secondary scintillation light signal (S2). The time delay between S1 and S2 signal encodes the depth of the interaction whereas the distribution of light on the top PMT array encodes the transverse position. The ratio between the S1 and S2 energy provides the discrimination between electronic and nuclear recoils, lastly the S1 and S2 energy provides information on the recoil energy. Figure reproduced from \cite{lutz2018} with permission.}
        \label{fig:tpc_wp}
    \end{center}
\end{figure}

The XENON1T TPC is suspended in a cryostat filled with 3.2 t of ultra-pure liquid xenon, 2 t of which is in the sensitive region of the TPC where particles scattering on the xenon can be detected and their position and energy reconstructed. The TPC is of cylindrical shape with a height and diameter of about 1 m. The TPC also consists of a field cage and various electrodes such that an electric field can be created inside the TPC to drift charges to the top (gaseous region) of the TPC. Here a second stronger electric field extracts the electrons from the liquid into the gaseous xenon where the electrons generate another scintillation light signal from their interactions with the xenon gas.
To detect both the prompt scintillation light (S1) from the initial scatter as well as the S2 from the electrons extracted at the top of the TPC, the TPC is instrumented with 248 photomultiplier tubes (PMTs). These circular 3-inch diameter Hamamatsu PMTs are placed in two arrays at the top and at the bottom of the TPC. The top array contains 127 PMTs, which are positioned in concentric rings in order to maximize the radial resolution of the position reconstruction.
Figure~\ref{fig:tpc_wp} shows the detector physics of a typical interaction in a dual--phase TPC. Further details about the XENON1T TPC can be found in \citep{aprile2017xenon1t}.

\begin{figure}[htbp]
\begin{center}
\includegraphics[width=2.5in]{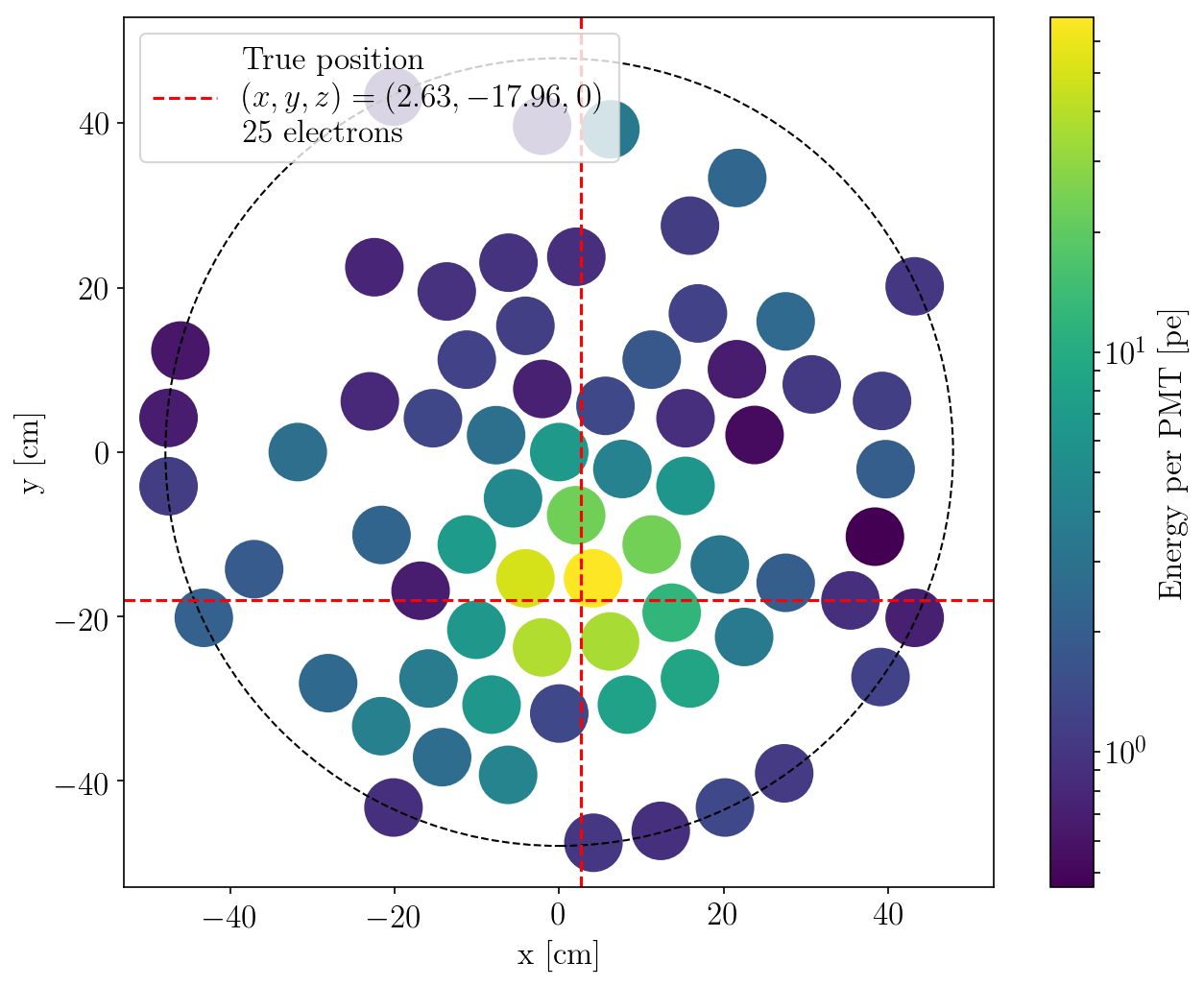}
\caption{The S2 hit pattern resulting from the secondary oscillation light S2 given the input coordinates $(x,y,z)=(2.63 , -17.96,0)$. The size of the charge signal of the event ($e$) is fixed equal to $25$.}
    \label{fig.topHP}
\end{center}
\end{figure}

The S2 signal is delayed by the drift time of the electrons through the TPC and this drift time is proportional to the depth at which the interaction occurred. This proportionality makes the drift time a very accurate estimator of the depth of the interaction $z$. In XENON100 the resolution of the vertical position of the interaction was found to be a factor $10$ better than the resolution of the ($x,y$) coordinates \citep{pelssers2015position}.
Since the S2 signal is always generated at the top of the TPC, very close to the top PMT array, the distribution of light over the top PMTs (the S2 hit pattern) contains information about the location of the S2 signal. Figure \ref{fig.topHP} shows the resulting S2 hit pattern for an event having input coordinates $(x,y,z)=(2.63\text{ cm} , -17.96\text{ cm}, 0\text{ cm})$.

\subsection{Position Reconstruction Algorithms in XENON1T}\label{posrecalg}
Several methods have been developed to reconstruct the 2--D (x,y) position using a S2 signal~\citep{Solovov:2011aa, Akerib:2017riv, pelssers2015position}. Next, we briefly describe some of the main algorithms provided in PAX: the ``maximum PMT'' method, a likelihood based method called Top Pattern Fit (TPF) and a method that uses an artificial neural network (NN). All these algorithms are implemented in PAX, the open source event reconstruction and data processing software developed by the XENON collaboration~\citep{xenon_collaboration_2018_1195785}. The version used in the following analyses is PAX v6.8.0, the same version that was used for the most recent XENON1T results \citep{aprile2018dark}.

The fastest and simplest method to reconstruct the 2--D (x,y) position, the ``maximum PMT'' method, looks for the PMT that captures the largest count of photoelectrons among the top PMT array, and takes that PMT and their corresponding 2--D coordinates as the position from which the S2 signal originated. This method has an uncertainty on the order of the distance between the top PMTs. The ``maximum PMT'' method can be used to give a rough estimate of the position of the S2 signal. We will however employ it in our first introductory example in Section~\ref{ex.fm1}, when using a simplified model.


The TPF method consists of an algorithm that, given a S2 hit pattern, returns the most likely 2--D (x,y) position, estimated by optimizing a likelihood function. The employed likelihood function is the Poisson likelihood chi--square $\chi_\lambda^2$ \citep{likelihood_chisquare}. This likelihood function assumes that the number of photoelectrons counted in a certain PMT is distributed according to a Poisson distribution with mean:
\begin{equation}
    \lambda_i = N_{obs} \frac{LCE_i(x, y)}{\sum_{j \in PMTs} LCE_j(x, y)},
\end{equation}
where $N_{obs}$ is the total number of observed photoelectrons in the S2 hit pattern, $LCE_i(x,y)$ is the light collection efficiency (LCE) of PMT $i$ for photons produced at position (x,y) and the sum is taken over all working PMTs in the top PMT array.
The LCE functions are not analytically known but are rather numerically estimated using optical photon Monte Carlo simulations. Those simulations take into account both the geometry of the detector and the optical and reflective properties of the employed materials. The LCE maps are not defined on the continuum but rather simulated on a grid, after which an interpolation is used.
In practice, the LCE maps evaluate the probability for a photon to reach a certain PMT given a set of assumed known 2--D position (x,y). For each live PMT on the top PMT array, a corresponding LCE map is implemented. By sampling from the LCE maps, an expected S2 hit pattern that originated from known 2--D positions (x,y) can be simulated, allowing to use the LCE maps to reconstruct the unknown 2--D positions (x,y) of events. Besides the most likely position, the TPF algorithm also returns the likelihood function values, defining the goodness--of--fit measure of the reconstructed S2 hit pattern. The method can also be used for retrieving confidence regions, since the likelihood surface is (partially) evaluated during the reconstruction.
The TPF method was the main position reconstruction algorithm used in \citep{aprile2017first}.

The last position reconstruction method is an artificial neural network (NN), trained with simulated S2 hit patterns from the same optical photon Monte Carlo simulation used to retrieve the LCE functions \citep{aprile2018dark}. NN provided the best resolution in XENON100 ($3$ mm) among all the methods used to reconstruct the 2--D (x,y) position at that time (TPF was introduced after XENON100). The performance of NN strongly depends on the resulting training set, and while the TPF method is physically motivated, the trained NN method learns to approximate the likelihood in an ad--hoc way. The network consists of a fully connected feed--forward network with 2 hidden layers, one input node for each PMT and two output nodes for (x,y). The NN method was the main position reconstruction algorithm used in \citep{aprile2018dark}.

In addition to the algorithms discussed here, several other algorithms such as the centroid algorithm~\citep{Solovov:2011aa} are also implemented in PAX. For our study we chose to compare against NN and TPF since those were found to perform best in the XENON1T detector.

In the following Sections of the present work we focus on a novel method, based on the likelihood--free framework, to reconstruct the  2--D (x,y) position. Beyond providing pointwise estimates, our method automatically provides credible regions for the parameters of interest. To retrieve a measure of the uncertainties related to each parameter of interest is crucial in order to discard events belonging to the background of the TPC. A key difference with respect to the TPF method is that with our proposed algorithm no LCE maps are needed as input to the optimization algorithm. Finally, one of the advantages of the proposed method lies in the fact that its procedure can be easily extended to more than two parameters. We show this feature in the last example where, together with the 2--D (x,y) position, also the size of the charge signal ($e$), a proxy for energy of the interaction, will be reconstructed.

\section{Statistical Methodology}\label{stat.method}

In this Section~\ref{stat.method} we introduce the likelihood--free framework, motivating its use and explaining the reasons for its success. In Section~\ref{abc} the Approximate Bayesian Computation (ABC) and the ABC--Population Monte Carlo (ABC--PMC) algorithms are presented, while in Section~\ref{bolfi} we introduce the Bayesian Optimization for Likelihood--Free Inference (BOLFI), a tool for accelerating likelihood--free inference that will be largely used in the examples presented in Section~\ref{examples}. Our choices for executing likelihood-free inference are discussed in Section~\ref{bolfi.TPC}.

\subsection{Approximate Bayesian Computation}\label{abc}

Bayesian inference has become an increasingly popular alternative to the frequentist approach over the last 20 years, thanks to several algorithmic advances that allow complex models to be fitted to data. In the frequentist framework the relation between observed data $r\obs$ and the vector of the parameters $\theta \in \Theta \subseteq  \mathbb R^p$ (where $p \ge 1$ is the dimension of the parameter space) can be fully described by the likelihood function $f(r\obs \mid \theta)$. In the Bayesian framework a prior distribution is assigned to the vector of parameters, $\theta \sim \pi(\theta)$. A Bayesian analyses is based on the so--called posterior distribution for $\theta$, defined according to the Bayes theorem:
\begin{equation}
\pi(\theta \mid r\obs) = \frac{f(r\obs \mid \theta)\pi(\theta)}{\int_{\Theta} f(r\obs \mid \theta)\pi(\theta) d\theta}, \label{posterior}
\end{equation}
where the denominator of Equation (\ref{posterior}) is a normalizing constant, often referred to as the evidence in computer science literature. The evaluation of Equation \eqref{posterior} relies on the ability to calculate the normalizing constant, but since in the vast majority of cases this quantity cannot be analytically calculated, the posterior is usually approximated using various sampling techniques such as Markov Chain Monte--Carlo (MCMC) algorithms \citep{hastings1970monte, metropolis1953equation}. As long as the likelihood is known, MCMC techniques are feasible. However this might not be the case, for example when the relationship between the data and the parameters is highly complex or unknown or if there are observational limitations. ABC is a framework of statistical inference designed for situations in which the likelihood function is intractable, but simulation through the forward model is possible. Recently, ABC has been applied in many different fields of science, such as astronomy \citep{akeret2015approximate, bruderer2016calibrated,CameronPettitt2012, hahn2017approximate,jennings2017astroabc, jennings2016new, WeyantEtAl2013,ShaferFreeman2012}, biology and epidemiology \citep{McKinleyEtAl2009}, ecology \citep{Beaumont2010}, population genetic problems \citep{beaumont2002approximate,cornuet2008inferring,PitchardEtAl1999,TavareEtAl1997} and population modeling \citep{TonyEtAl}.  
%
%

The resulting ABC posterior distribution, following the notation from \cite{MarinEtAl2012}, can be written as:
\begin{equation}
\pi_{\epsilon}(\theta \mid s(r\obs)) = \int \left[ \frac{f(s(r\pro) \mid \theta)\pi(\theta)\mathbb I_{A_{\epsilon,s(r\obs)}}(s(r\pro))}{\int_{A_{\epsilon,s(r\obs)} \times \Theta} f(s(r\pro) \mid \theta)\pi(\theta) d s(r\pro) d\theta}\right]ds(r\pro), \label{ABCposterior}
\end{equation}
where $\mathbb I_{A_{\epsilon,s(r\obs)}}(\cdot)$ is the indicator function for the set $A_{\epsilon,s(r\obs)} = \{s(r\pro) \mid \rho(s(r\obs), s(r\pro)) \leq \epsilon\}$.

According to the definitions provided through Equation (\ref{posterior}) and Equation~\eqref{ABCposterior} for the true posterior distribution and the ABC posterior distribution respectively, it follows that $\pi(\theta \mid r\obs) \approx \pi_{\epsilon}(\theta \mid s(r\obs))$ for $\epsilon \to 0$ and $s(\cdot)$ sufficient. In practice, however, for computational reasons, some tolerance $\epsilon > 0$ has to be allowed, which causes an approximation error in the ABC procedure. Secondly, rather than comparing the entire observed data $r\obs$ with the simulated sample $r\pro$, the similarity between the observed and the simulated data is based on suitably selected summary statistics. While a desirable situation involves selecting summary statistics $s(\cdot)$ that are sufficient \citep{cox1979theoretical}, this rarely happens when facing real problems necessitating ABC. As pointed out in \cite{MarinEtAl2012}, for most situations the summary statistics are usually determined by the problem at hand and selected by the experimenters in the field, making the implementation of a general procedure for retrieving highly informative summary statistics challenging. Provided that the summary statistics and the tolerance are properly selected, the ABC posterior distribution suitably matches the true posterior distribution \citep{beaumont2002approximate,BlumEtAl2013,blum2010choosing,FearnheadPrangle2012,IshidaEtAl2015,prangle2015summary}. 
The original basic ABC algorithm \citep{PitchardEtAl1999, TavareEtAl1997} can require a huge computational time in order to produce the desired number of posterior samples, because it only uses prior distributions throughout the procedure \citep{MarinEtAl2012}. Recently, many algorithms that extend the basic ABC algorithm have been proposed \citep{Blum2010,BlumEtAl2013,CsilleryEtAl2010,DrovandiEtAl2011,FearnheadPrangle2012,gutmann2016bayesian,JoyceMarjoram2008,MarinEtAl2012,DelMoralEtAl2012,RatmannEtAl2013}, but at least for the first example presented in this work we focus on the ABC--PMC algorithm \citep{BeaumontEtAl2009}. 

The ABC--PMC algorithm, in order to improve the efficiency of the statistical inference, constructs a series of intermediate distributions rather than only using the prior distributions. The first iteration of the ABC--PMC algorithm uses tolerance $\epsilon_1$ and draws proposals from the specified prior distribution.
Starting from the second iteration, the algorithm draws proposals from the previous iteration's ABC posterior distribution. After a particle is selected from the previous iterations particle system, it is moved according to some kernel. Since the obtained proposals are not directly drawn from the prior distributions, importance weights are used.  The importance weight for particle $J = 1, \ldots, N$ at iteration $t$ is defined as:

\begin{equation*}
W_t^{(J)} \propto \pi(\theta_{t}^{(J)})/\sum_{K = 1}^N W_{t-1}^{(K)} \phi\left[\tau_{t-1}^{-1}\left(\theta_{t}^{(J)} - \theta_{t-1}^{(K)}\right)\right],
\end{equation*}
where $\phi(\cdot)$ is a Gaussian kernel with variance $\tau_{t-1}^2$ (i.e. twice the weighted sample variance of the particles from iteration $t-1$, as originally suggested by \cite{BeaumontEtAl2009}). The steps required to run the ABC--PMC algorithm are displayed in Algorithm~\ref{alg2:ABC--PMC.alg}.

\begin{algorithm}
\caption{ABC--PMC algorithm for $\theta$}
\begin{algorithmic}
\If{$t = 1$}
\For{$J = 1, \ldots, N$}
\State Set $d_1^{(J)} = \epsilon_1 + 1$
 \While{$d_1^{(J)} > \epsilon_1$}
	\State Propose $\theta^{(J)}$ by drawing $\theta\pro \sim \pi(\theta)$,
 	\State Generate  $r\pro \sim f\left(\cdot \mid \theta^{(J)}\right)$
 	\State Calculate distance $d_1^{(J)} = \rho(s(r\obs), s(r\pro))$
 \EndWhile
 	\State Set weight $W_1^{(J)} = N^{-1}$
 \EndFor
 \ElsIf{$2 \leq t \leq T$}
  \State Set $\tau_{t}^{2} = 2 \cdot \text{var} \left( \{\theta_{t-1}^{(J)},W_{t-1}^{(J)}\}_{J=1}^{N}\right)$
 \For{$J = 1, \ldots, N$}
 \State Set $\epsilon_t = q^{th}$ quantile of $\{d_{t-1}^{(J)}\}_{J = 1}^N$
 \State Set $d_t^{(J)} = \epsilon_t + 1$
 \While{$d_t^{(J)}> \epsilon_t$}
	\State Select $\theta_t^*$ from $\theta_{t-1}^{(J)}$ with probabilities $\left\{W_{t-1}^{(J)}/\sum_{K = 1}^NW_{t-1}^{(K)}\right\}_{J = 1}^N$
	\State Propose $\theta_t^{(J)} \sim \mathcal{N}(\theta_t^*, \tau_{t}^{2})$
	\State Generate  $r\pro \sim f\left(\cdot \mid \theta_{t}^{(J)}\right)$
 	\State Calculate distance $d_t^{(J)} = \rho(s(r\obs), s(r\pro))$
 \EndWhile
	\State Set weight $W_t^{(J)} \propto \pi(\theta_t^{(J)})/\sum_{K = 1}^N W_{t-1}^{(K)} \phi\left[\tau_{t-1}^{-1}\left(\theta_{t}^{(J)} - \theta_{t-1}^{(K)}\right)\right]$
 \EndFor
 \EndIf
\end{algorithmic} \label{alg2:ABC--PMC.alg}
\end{algorithm}

The series of tolerances decreases such that $\epsilon_1 > \epsilon_2 > \cdots > \epsilon_T$, where $T$ is the final iteration of the ABC--PMC algorithm. Both the rule to reduce the tolerances and the total number of iterations $T$ are selected in advance and must be properly tuned for the considered application \citep{BeaumontEtAl2009, IshidaEtAl2015, Lenormand2013, McKinleyEtAl2009,SissonEtAl2007,TonyEtAl,WeyantEtAl2013}. We will discuss our choices to run the ABC--PMC algorithm in Section~\ref{examples}. The result of the ABC--PMC analysis consists of the approximate posterior distribution, used to provide both pointwise estimates and credible regions on the parameters of interest. 

One of the advantages of working in the Bayesian framework is the availability for calculating credible intervals starting from the posterior distribution. Given the general posterior distribution $\pi(\theta \mid r\obs)$, $B$ is a credible set for $\theta$ if:
\begin{equation}
\text{Pr}(\theta \in B | r\obs) = \int_B \pi(\theta \mid r\obs) d\theta.
\label{eq:credible.interval}
\end{equation}
As an example, a credible interval for $\theta$ of level $100 (1 - \alpha) \%$, where $\alpha$ is the first type error, is defined as that interval $B$ such that $\int_B \pi(\theta \mid r\obs) d\theta = 0.95$ \cite{eberly2003estimating}. Among the others, an often used credible interval is the Highest Posterior Density (HPD) interval \citep{chen1999monte, sinha1983credible}. For a general HPD interval of level $\alpha$, the posterior probability for the region $B$ is $100 (1 - \alpha) \%$. Moreover, the minimum density of any point within the region $B$ is equal to or larger than the density of any point outside that region. An HPD is that interval for which most of the distribution lies \cite{turkkan1993computation}. We stress on the fact that one of the advantages of the proposed algorithm is the availability non only for pointwise estimates but also for credible intervals for the parameters of interest.

\subsection{Bayesian Optimization for Likelihood--Free Inference}\label{bolfi}

One of the major obstacles to likelihood--free inference is the computational cost of the method. In fact ABC is based on the idea of identifying relevant regions of the parameters of interest of the model by finding those estimates such that $r\pro$, or more often its summary $s(r\pro)$, is comparable with $r\obs$, or more often its summary $s(r\obs)$. It is clear then that the role played by the summary statistics and by the discrepancy measure defined to compare $s(r\pro)$ and $s(r\obs)$ is key in order to retrieve useful statistical properties, and in particular making the ABC posterior distribution a suitable approximation of the true posterior. Most of the parameters proposed during the resampling step result in large distances between $s(r\pro)$ and $s(r\obs)$ and those estimates for the parameters for which the distances would be small are unknown, as pointed out in \cite{lintusaari2017fundamentals}. This latter fact means that, when using the ABC--PMC algorithm presented in Section~\ref{abc}, the number of datasets to simulate through the forward model is usually at least on the order of $10^6$, since weak information is generally available in advance about relevant regions of the parameter space.

In order to address the issue related to the decision about the similarity between $s(r\pro)$ and $s(r\obs)$, an alternative to the first class of algorithms able to reduce the computational burden by several order of magnitude is provided by BOLFI \citep{gutmann2016bayesian}. BOLFI combines probabilistic modelling of the distance function that compares $s(r\pro)$ with $s(r\obs)$ via optimization, to facilitate likelihood-free inference. The basic idea behind BOLFI is to find, avoiding unnecessary computations, relevant regions of the parameter space where the discrepancy between $s(r\pro)$  and $s(r\obs)$  is small. This is done by using a probabilistic model to learn on the stochastic relation between the parameter estimates and the similarity between $s(r\pro)$  and $s(r\obs)$ . Once the probabilistic model is ready, it can be used to retrieve a suitable approximation to the true posterior distribution \citep{gutmann2016bayesian}. Therefore, the problem becomes to infer the regression function of the discrepancies, which is unknown, focusing on regions of the parameter space where those discrepancies are small. In their work, \cite{gutmann2016bayesian} proposed to use Bayesian optimization \citep{brochu2010tutorial}, modelling the function of the discrepancies with a Gaussian process. The choice for using Gaussian processes to model the function of the discrepancies is not mandatory. Recently Gaussian processes have been used as surrogate models for evaluating generative models that are expensive to compute \cite{meeds2014gps,wilkinson2014accelerating,rasmussen2004gaussian}. In the present work, we followed the specifications suggested by \citep{gutmann2016bayesian}. Further details on BOLFI can be found in \cite{gutmann2016bayesian} and the Python package called ELFI \citep{ELFI} is freely available for likelihood--free inference, offering both the ABC--PMC algorithm and BOLFI as available inference options.

\subsection{Likelihood--free inference for event reconstruction in a dual-phase TPC}\label{bolfi.TPC}

As highlighted in the previous Sections, one of the most important choices when working in the likelihood--free framework is the selection of summary statistics highly informative on the parameters of interest. Since the comparison between the entire simulated S2 hit pattern, $r\pro$, and the entire observed S2 hit pattern, $r\obs$, is computationally unfeasible, reducing the sample space without losing information on the  2--D (x,y) position is crucial. For the examples presented in Section \ref{examples}, we generally perform two separated ABC--PMC/BOLFI analyses. The first analysis uses as summary statistic the well known Euclidean distance while for the second analysis the Bray--Curtis dissimilarity is used. The Bray--Curtis dissimilarity is not a true metric, since it does not have the triangle inequality property. Despite this, for our goal of properly comparing $r\obs$ with $r\pro$, we found the Bray--Curtis extremely useful. One of the advantages of the Bray--Curtis relies on its interpretability: a Bray--Curtis dissimilarity equal to $0$ means that $r\obs$ and $r\pro$ are exactly the same, while a value equal to $100$ defines the maximum difference that can be observed between two S2 hit patterns. Moreover the Bray--Curtis dissimilarity, originally used in ecology \citep{clarke2006resemblance}, works under the assumption that the samples are taken from the same physical size, such that for instance the PMTs that count the number of photoelectrons. Given the observed S2 hit pattern, $r\obs$, and a simulated one, $r\pro$, the summary statistics based on the Euclidean distance and the Bray--Curtis dissimilarity are respectively defined as:

\begin{equation}
\rho(r\obs, r\pro)_{\text{Euclidean}} = \sqrt{\sum_{i=1}^{n} (r\obs^i - r\pro^i)^2}
\label{eq.summarystat}
\end{equation}

and

\begin{equation}
\rho(r\obs, r\pro)_{\text{Bray--Curtis}} = \frac{\sum_{i=1}^{n} | r\obs^i - r\pro^i |}{\sum_{i=1}^{n} | r\obs^i + r\pro^i |},
\label{eq.summarystat.bc}
\end{equation}
where $n$ is the total number of PMTs composing the S2 hit pattern, assuming all the PMTs are operational. By using Equation \eqref{eq.summarystat} or Equation \eqref{eq.summarystat.bc} we reduce the sample space from 128 to 1, while preserving the information carried out by the entire sample, as will be shown in the coming examples. We note that for the third and last example a further distance, namely the Energy distance, will be as well used and combined with the Bray--Curtis dissimilarity in order to reconstruct the 3--D coordinates ($x,y,e$). Details are found in Section \ref{ex.fm3}.

We end this Section by presenting the statistical tests used to compare the performances of the likelihood--free algorithm and the already available options, focusing in particular on TPF and NN. Using large-scale simulation studies, for each method and given some known inputs, the 2--D (x,y) position or the 3--D coordinates ($x,y,e$) of an event is reconstructed. Then, we calculated the Euclidean distance from the input position for each of the used methods. In particular, for the examples presented in Section \ref{ex.fm1} -- \ref{ex.fm2} that reconstruct the 2--D (x,y) position, defined $(x_\text{input}, y_\text{input})$ and $(x_\text{rec}, y_\text{rec})$ as respectively the input coordinates and the reconstructed coordinates with one of the different methods, the Euclidean distance $d_{\text{euc}}$ in $\mathbb{R}^2$ is defined as:
\begin{equation}
d_{\text{euc}} = \sqrt{(x_\text{input}-x_\text{rec})^2 + (y_\text{input}-y_\text{rec})^2}.
\label{eq.eucXY}
\end{equation}

To quantify the goodness of the 3--D reconstruction presented through the third example of Section \ref{ex.fm3}, the Euclidean distance $d_{\text{euc}}$ in $\mathbb{R}^3$ between the reconstructed event $(x_\text{rec}, y_\text{rec}, e_\text{rec})$ and the input coordinates $(x_\text{input}, y_\text{input}, e_\text{input})$ is retrieved, according to Equation \eqref{eq.eucXYE}:
\begin{equation}
d_{\text{euc}} = \sqrt{(x_\text{input} - x_\text{rec})^2 + (y_\text{input} - y_\text{rec})^2 + (e_\text{input} - e_\text{rec})^2}.
\label{eq.eucXYE}
\end{equation}

We note that since the 2--D (x,y) position and the charge signal $e$ have different units of measure, Equation \eqref{eq.eucXY} and Equation \eqref{eq.eucXYE} define distances on different parameter spaces. In particular, we defined the range of the charge signal to be the same order of magnitude than the 2--D (x,y) positions, as shown in Section \ref{ex.fm3}.
Finally, statistical tests are performed to compare the results obtained by BOLFI with TPF, the latter being the best currently existing method among the available ones. Because the different reconstruction algorithms are used on the same set of events, the most suitable statistical test would be the paired sample t test \citep{mee1991regression}. Unfortunately, for all the considered examples of Section \ref{examples}, the assumption of normality underlying the paired sample t test does not hold. In fact the Shapiro--Wilk test used to test the normality assumption of the distribution provides P--values for which the hypothesis of normality is always strongly rejected. For this reason we used the non parametric Wilcoxon Signed Rank Test \citep{wilcoxon1970critical}.

\section{Examples and discussions}\label{examples}

In this Section we show the advantages of using the likelihood--free framework to reconstruct the 2--D (x,y) position over the currently existing alternative methods quickly reviewed in Section \ref{posrecalg}. We do so by presenting three numerical examples. In the first example we use a simple forward model in order to reconstruct the 2--D (x,y) position starting from a S2 hit pattern. The second and the third example use a more complex and realistic forward model. Therefore, for computational reasons, BOLFI is required to perform the likelihood--free analyses. In the second example we are interested just in reconstructing the  2--D (x,y) position of an event, while in the third and final example we add a third variable, the size of the charge signal ($e$), making a three parameter inference problem (x,y,e). The softwares used to perform the analyses and to produce the Figures are Python\footnote{http://www.python.org} and \textsf{R}.\footnote{https://cran.r-project.org}

\subsection{A first simple simulation example}\label{ex.fm1}

In this first example we start investigating the feasibility for reconstructing the  2--D (x,y) position given the S2 hit pattern by using the likelihood--free framework and in particular by implementing the ABC--PMC algorithm introduced in Section~\ref{abc}. In order to do so, we use a simple forward model that samples from the LCE maps. This forward model, that returns for a given set of 2--D coordinates (x,y) the corresponding S2 hit pattern, requires to specify the LCE map binning (zoom factor) and the number of photons to be specified. Since we use the LCE maps directly the spatial resolution will be limited by the grid spacing of the LCE maps. To ensure the limiting factor is due to the binning of the LCE maps and not due to our algorithm we fix the zoom factor equal to twice the LCE map zoom factor (giving us twice the resolution). We fix the number of detected photons to $500$. Given the discrete nature of this simple forward model, we add stochastic noise from a standard Normal distribution to the resulting S2 hit pattern.

Beyond the definition of the forward model, prior distributions have to be assigned to the parameters of interest, that are in this case the 2--D coordinates x and y. For (x,y), according to Equation \eqref{eq.priorABC}, a Bivariate Normal prior distribution is used, where the means are the coordinates ($x_\text{max PMT}, y_\text{max PMT}$) estimated by the ``maximum PMT'' method and the covariance matrix is diagonal with standard deviations equal to $15$ cm. The parameters previously defined are also known as the hyper--parameters of the prior distribution. We recall that the ``maximum PMT'' method estimates as reconstructed position that PMT with the largest count of photonelectrons.
\begin{equation}
\begin{pmatrix}
x_\text{prop} \\
y_\text{prop}
\end{pmatrix} \sim  N_{2}
\begin{bmatrix}
\begin{pmatrix}
x_\text{max PMT}\\
y_\text{max PMT}
\end{pmatrix}\!\!,&
\begin{pmatrix}
15 & 0 \\
0 & 15
\end{pmatrix}
\end{bmatrix}. \\[2\jot]
\label{eq.priorABC}
\end{equation}
A coordinate $(x_\text{prop}, y_\text{prop})$ is valid if the following constraint, related to the dimension of the TPC, is satisfied:
\begin{equation}
    x_\text{prop}^2+y_\text{prop}^2 \le 47.9^2 \text{cm}^2.
\label{eq.priorABC.constraint}
\end{equation}
Once the constraint is satisfied, $(x_\text{prop}, y_\text{prop})$ can be used to obtain the simulated S2 hit pattern, $r\pro$, and the steps outlined in Algorithm \ref{alg2:ABC--PMC.alg} can take place.

The last required step before running the ABC--PMC algorithm consists of tuning some of the parameters originally introduced by \cite{BeaumontEtAl2009}. In particular four parameters need to be properly selected: the desired particle sample size $N$, the total number of iterations $T$, the initial tolerance $\epsilon_1$ and finally the rule to reduce the tolerance through the iterations. These parameters are not tuned once, but they are rather the result of successive attempts and adjustments that combine computational savings with a suitable approximation by the ABC posterior distribution to the true posterior. We defined the desired particle sample size $N=1000$, the total number of iterations $T=40$, $\epsilon_1 = 4$ and finally we adaptively selected the next tolerance of the algorithm by calculating the $q_{\text{t}}=85^{th}$ percentile of $\{d_{t-1}^{(J)}\}_{J = 1}^N$, the distances of the previous step accepted particles system. There are of course other possible choices: when choosing $T$ and $q_t$, the most important aspect to consider is to avoid stopping the algorithm too soon (i.e. the ABC posterior distribution is a poor approximation of the true posterior) as well as to avoid to run the algorithm for too long (i.e. the overall efficiency of the algorithm is low). We found that, with the chosen quantile, after 40 iterations further reductions of the tolerance did not lead to a better approximation of the ABC posterior distribution, suggesting that a suitable approximation to the true posterior is reached. As outlined in Algorithm \ref{alg2:ABC--PMC.alg}, starting from the second iteration, rather than using the prior distribution, a perturbation kernel is used. We used the Gaussian perturbation kernel proposed in \cite{BeaumontEtAl2009}, taking into account the previously defined constraint on the parameters and calculating the importance weights accordingly.

\begin{figure}[htbp]
\begin{center}
\includegraphics[width=3.5in]{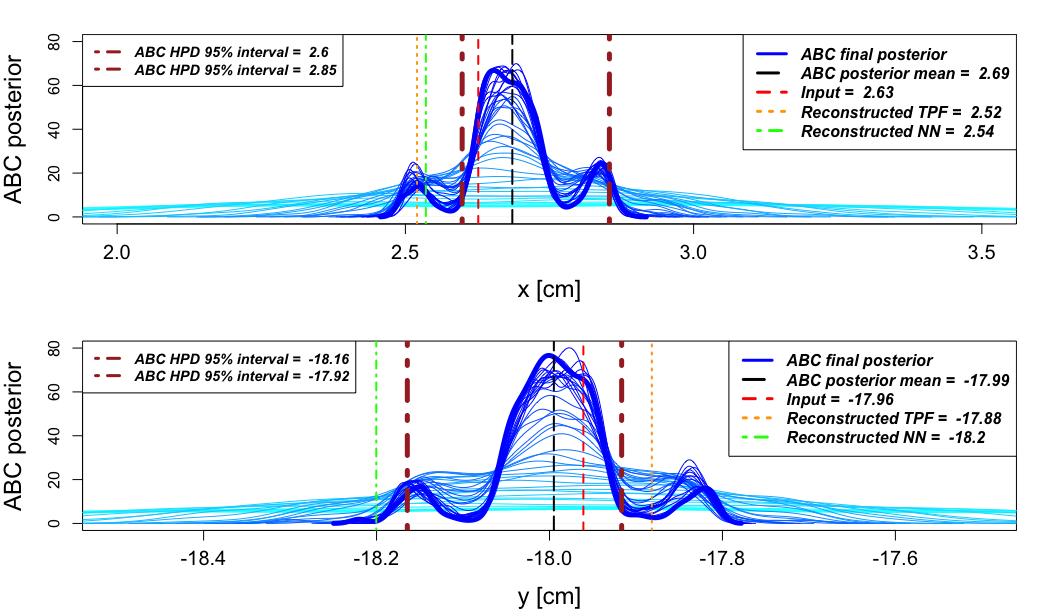}
\includegraphics[width=2.3in]{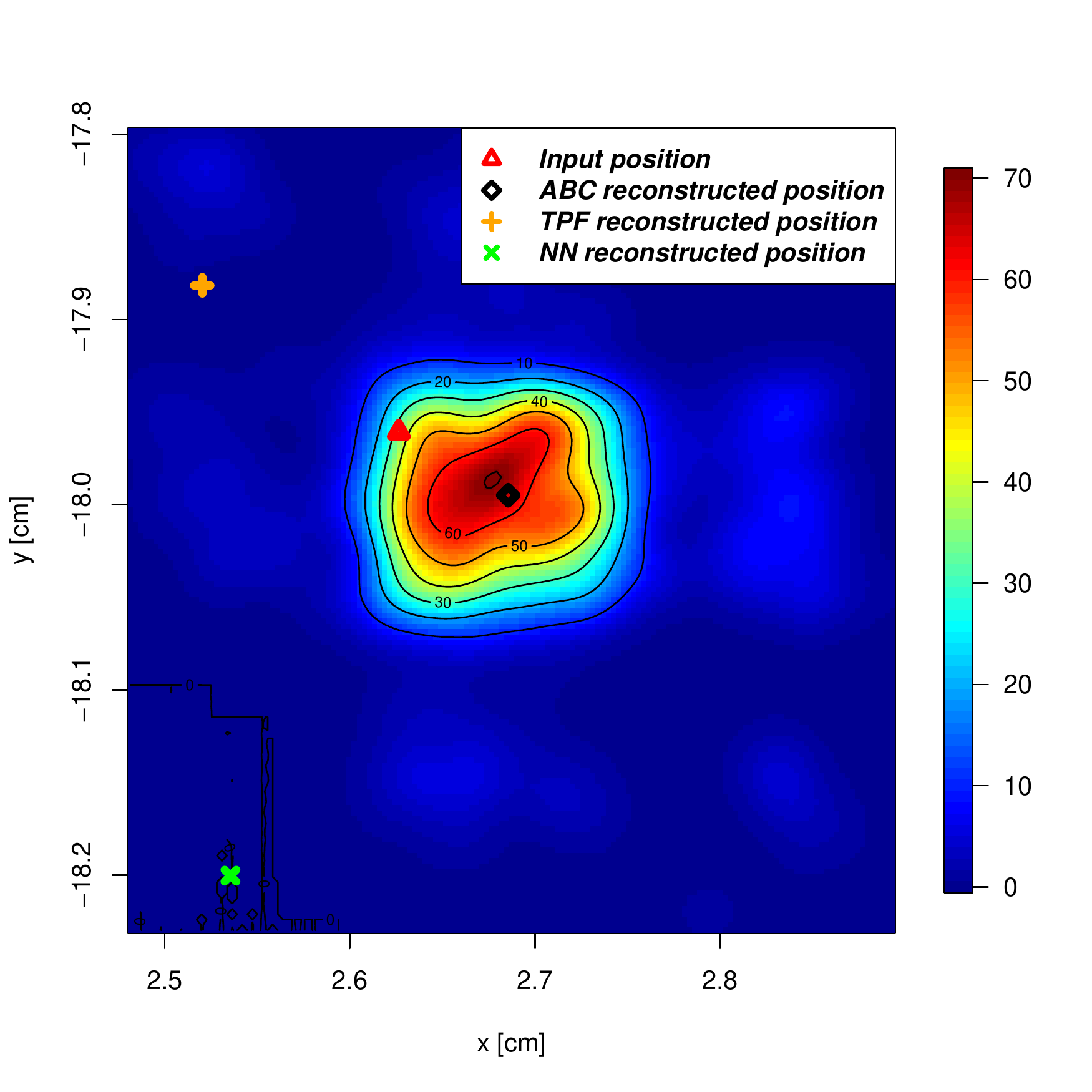}
\caption{(left) ABC posterior distributions for x (top) and y (bottom) estimated at the end of each iteration. At the beginning of the procedure (cyan solid lines) the ABC posterior distributions are broader because of the large used tolerance $\epsilon$. Once $\epsilon$ is sufficiently small, the ABC posterior distribution stabilizes and further reductions of the tolerance do not improve the posterior (bluer solid lines). The final tolerance is $\epsilon_{40}=0.76$. For the final ABC posterior distributions, obtained after $T=40$ iterations, the corresponding HPD 95\% interval are displayed for both x and y(brown dashed--point lines). (right) Bivariate contour plot of the joint ABC posterior distribution (x,y) and its corresponding pointwise estimates, obtained with ABC--PMC (black square), TPF (orange cross) and NN (green x). Those reconstructed positions are compared with the original input (red triangle). The reconstructed positions obtained by TPF and NN are outside the 2--D surface of the posterior and do not belong to the HPD 95\% interval.}
    \label{fig.simpleFM}
\end{center}
\end{figure}

\begin{figure}[htbp]
\begin{center}
\includegraphics[width=4in]{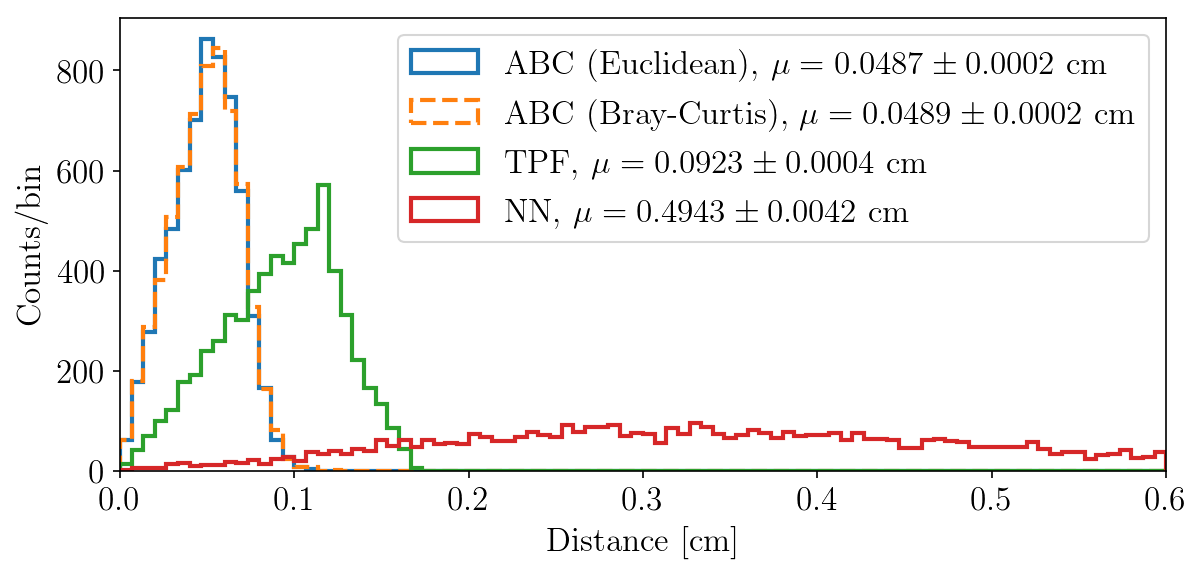}
\caption{Euclidean distances obtained using Equation \eqref{eq.eucXY} for all the four methods resulting from the reconstruction of 6297 independent events. The proposed ABC--PMC algorithm improves the reconstruction with respect to the currently existing alternative methods for both the selected summary statistics. The average Euclidean distances from the input for the positions reconstructed with ABC--PMC (and by using as summary statistic Equation \eqref{eq.summarystat}), ABC--PMC (and by using as summary statistic Equation \eqref{eq.summarystat.bc}), TPF and NN are respectively equal to $0.0487\text{ cm}$, $0.0489\text{ cm}$, $0.0923\text{ cm}$ and $0.4943\text{ cm}$. Note that the factor $2$ improvement between ABC--PMC and TPF is due to the same factor $2$ difference in the zoom factor. Since the likelihood is known for this forward model we would get the same performance from ABC--PMC and TPF for the same zoom factor. Using a higher zoom factor shows that the LCE map binning is indeed the limiting factor and not due to the ABC--PMC algorithm itself. Secondly we observe that for this forward model both summary statistics (Euclidean and Bray--Curtis) are equally informative.}
    \label{fig.simpleFMcomparison}
\end{center}
\end{figure}

Once all the necessary quantities are defined, the ABC--PMC algorithm can be executed in order to reconstruct the  2--D (x,y) position. As an example of that, we consider an event whose true position is $(x_\text{input}=2.63\text{ cm}, y_\text{input}=-17.96\text{ cm})$. The z coordinate is fixed equal to $0$. By entering $(x_\text{input}, y_\text{input})$ in the forward model we obtain the observed S2 hit pattern, $r\obs$. The goal is, using only $r\obs$, to reconstruct the position of the S2 event. The results of our ABC--PMC analysis are displayed in Figure \ref{fig.simpleFM} and summarized in Table \ref{table.simpleFM}. The summary statistic used to produce the present plots is the Euclidean distance defined in Equation \eqref{eq.summarystat}.
Looking at the left plot of Figure \ref{fig.simpleFM} we can evaluate the final ABC marginal posterior distributions, obtained after $T=40$ iterations and non--parametrically estimated, for x and y.
The ABC marginal posteriors exhibit two local modes apart from the global maximum. Note that both the TPF and the NN methods seem to find reconstructed position coordinates near the local modes.
Through the $40$ iterations, the tolerance $\epsilon$ decreases, allowing the procedure to move from poorly informative ABC posterior distributions (cyan solid lines, corresponding to the first iterations) to stable and informative ones (bluer solid lines), once the tolerance has sufficiently decreased. The final tolerance is $\epsilon_{40}=0.76$. Together with the posterior distributions, we used as pointwise statistic that defines the reconstructed coordinate the posterior mean. In this case, since the ABC--PMC uses importance weights, the posterior means for x and y are the result of weighted means. There are of course other possible choices when picking a pointwise indicator from a posterior distribution, such as the maximum at posteriori (MAP) estimate. We found for the present analyses the posterior mean to provide the closest results to the input coordinates. Once samples from the posterior distribution are available, it is possible to retrieve a credible interval for the parameters of interests. In this case we calculated  a HPD 95\% credible interval for both x and y. The HPD interval provides useful information: first of all we can see how these intervals contain the input coordinates. As second, the HPD interval can be used to evaluate the accuracy of the position reconstructed with other methods, in this case TPF and NN.
The right side plot of Figure \ref{fig.simpleFM} shows the bivariate contour plot, where it is possible to observe that, for the presented example, the positions reconstructed using ABC are much closer to the input coordinates that any reconstructed position obtained with the currently existing alternative methods. 
\begin{center}
\begin{table}[ht]
\centering
\resizebox{\columnwidth}{!}{%
\begin{tabular}{|c|c|c|c|c|c|}
\hline
 & Input Values &  ABC posterior means (HPD 95\% interval) & TPF method & NN method & maximum PMT method\tabularnewline
\hline
\hline
x [cm]& 2.63  & 2.69 (2.60; 2.85)  &  2.52  & 2.54  &  4.12 \tabularnewline
\hline
y [cm]& -17.96 & -17.99 (-18.16; -17.92)  & -17.88  & -18.20 & -15.36  \tabularnewline
\hline
\end{tabular}}
\caption{Comparison between the ABC posterior means, used as pointwise statistics, and the estimates obtained with the currently existing alternative methods. Together with the pointwise estimates for the ABC--PMC analysis, a HPD 95\% interval is displayed. The last column defines the reconstructed (x,y) position estimated with the ``maximum PMT'' method, employed to define the hyper--parameters for the prior distribution used in the ABC--PMC analyses.}
\label{table.simpleFM}
\end{table}
\end{center}
We performed the ABC--PMC positioning reconstruction analysis over 6297 independent events. Then, we calculated the Euclidean distance from the input position for each of the four used methods, according to Equation \eqref{eq.eucXY}. 
%
%
The results, displayed in Figure \ref{fig.simpleFMcomparison}, show that the proposed ABC--PMC algorithm improves the accuracy of the reconstruction with respect to the currently existing alternative methods both when using the Euclidean distance or the Bray--Curtis dissimilarity as summary statistic. The average Euclidean distance for the positions reconstructed with ABC--PMC that use as summary statistic to compare $r\pro$ and $r\obs$ Equation \eqref{eq.summarystat} is $0.0487\text{ cm}$. When using Equation \eqref{eq.summarystat.bc} as a summary statistic the average Euclidean distance is as well $0.0489\text{ cm}$. The average Euclidean distances for the positions reconstructed with TPF and NN are respectively equal to $0.0923\text{ cm}$ and $0.4943\text{ cm}$. In this simple initial example the smallest distances from the input coordinates are obtained when using the ABC--PMC algorithm and as summary statistic to compare $r\pro$ and $r\obs$ Equation \eqref{eq.summarystat} or Equation \eqref{eq.summarystat.bc}. The main explanation for why the ABC--PMC analysis outperformed the TPF method lies in the fact that, at least for this simple forward model, TPF gives discrete reconstructed positions, while ABC gives continuous ones. We note that the factor $2$ improvement between ABC--PMC and TPF is due to the same factor $2$ difference in the zoom factor. 

The goal for this first example that uses a simple forward model that samples from the LCE maps, and whose likelihood is tractable, was to introduce the ABC--PMC algorithm as a novel framework to reconstruct the 2--D ($x,y$) position using as information the S2 hit pattern. Since for this simple forward model TPF gives discrete reconstructed positions while ABC gives continuous ones, operating a comparison of the performances with respect to the currently existing alternative methods is unfeasible.  We tested the feasibility for a likelihood--free analysis, defined prior distributions for the parameters of interest and retrieved highly informative summary statistics. In the next two examples more complex forward models having intractable likelihoods will be used and also TPF will give continuous reconstructed positions, allowing for a comparison between our proposed method and the currently existing alternative methods. The full waveform simulation and BOLFI as the statistical tool for the likelihood--free inference will be used in order to execute the analyses.

\subsection{Position Reconstruction using full waveform simulation}\label{ex.fm2}

The first example introduced in Section~\ref{ex.fm1} is useful to show from a practical standpoint how to implement the ABC--PMC algorithm and to show the potential of the likelihood--free framework in reconstructing the  2--D (x,y) position. Aimed by the same goal, the second example presented below, rather than using a simple forward model that samples from the LCE maps, employs the complete waveform simulation and reconstruction. Once provided the input coordinates $(x_\text{input}, y_\text{input})$, beyond the S2 top hit pattern this forward model also returns the S2 bottom hit pattern and the time required by the photo--electrons to reach the top PMTs, known with the term timing. The forward model returns the S2 bottom hit pattern together with the S2 top hit pattern by considering the light distribution as well as the light collection efficiency of the S2 signal over all PMTs in the TPC. We note that in the previous example we called the S2 top hit pattern simply S2 hit pattern. From now on we will clearly distinguish the type of S2 hit pattern we are referring to, specifying the terms ``bottom'' or ``top''. Since TPF and NN only consider the S2 top hit pattern to reconstruct the  2--D (x,y) position, we consider for this second example only the S2 top hit pattern as well, in order to present a comparison that uses the same amount of information among the employed methods.

The new forward model is much more complex and realistic and therefore slower than the one used in Section~\ref{ex.fm1}, making the use of the ABC--PMC algorithm computationally unfeasible. For this reason, rather than the ABC--PMC algorithm, the likelihood--free analysis is carried out using BOLFI, introduced in Section~\ref{bolfi}. In the following we consider all the PMTs to be operational. We consider two different sizes of the charge signal, one releasing $10$ electrons, another one releasing $25$ electrons. As prior distribution for ($x_\text{prop},y_\text{prop}$) we use Equation \eqref{eq.priorABC}, but the mean hyper--parameters are taken from the TPF method rather than from the ``maximum PMT'' method. Also in this case, when using BOLFI to reconstruct the 2--D (x,y) position, we perform the analyses for two different summary statistics: in the first analysis we use the Euclidean distance defined through Equation \eqref{eq.summarystat}, while in the second analysis the Bray--Curtis dissimilarity defined through Equation \eqref{eq.summarystat.bc} is used.

\begin{figure}[htbp]
\begin{center}
\includegraphics[width=5in]{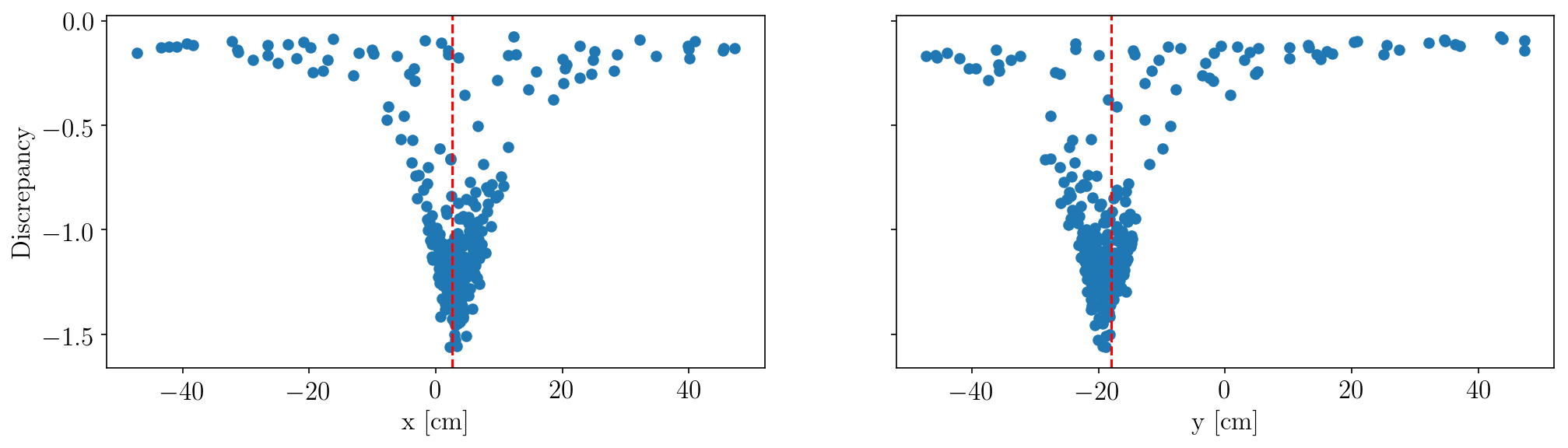}
\includegraphics[width=2.5in]{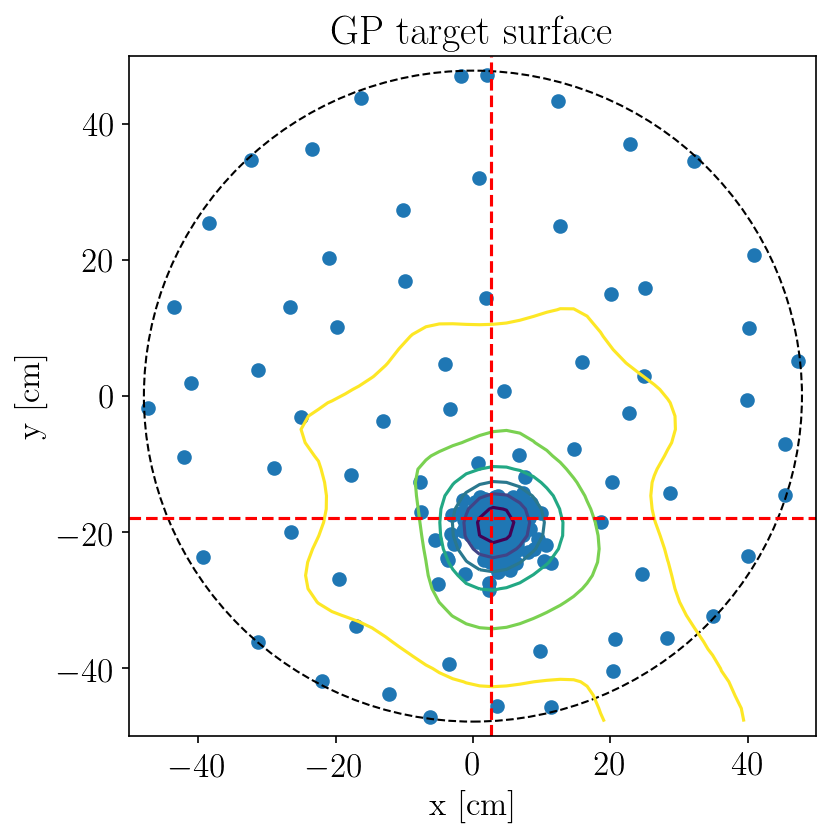}
\includegraphics[width=2.5in]{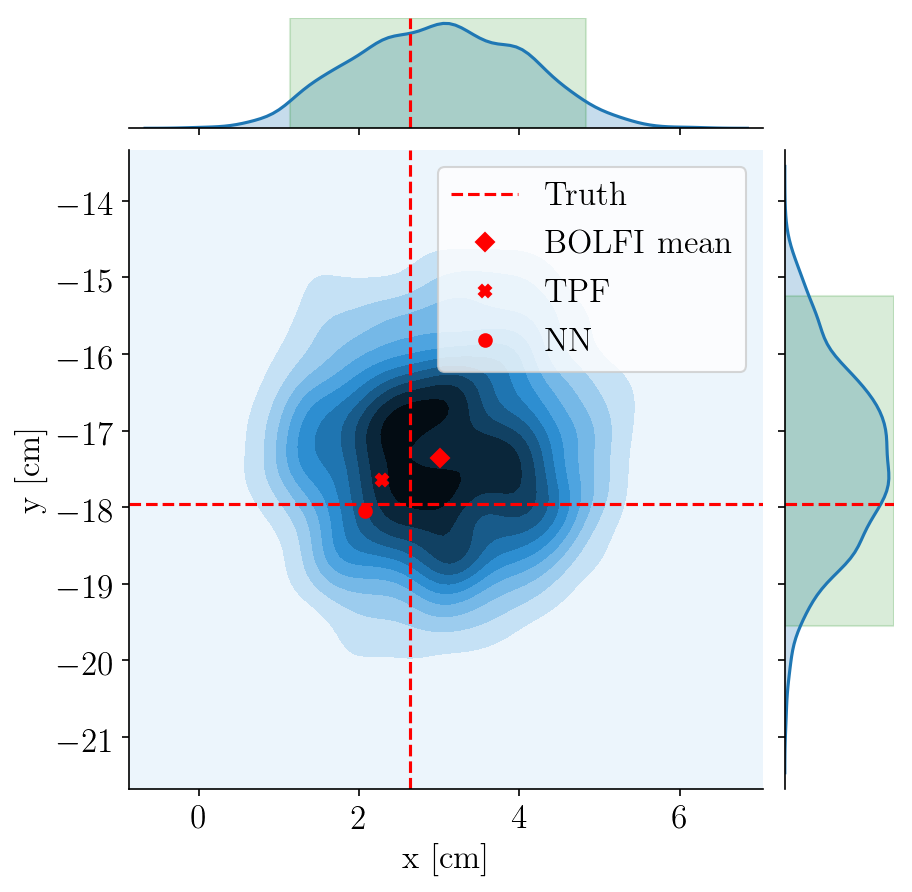}
\caption{(top) Logarithm of the discrepancies returned by BOLFI for a generic  2--D (x,y) position, using as summary statistic the Bray--Curtis dissimilarity defined in Equation \eqref{eq.summarystat.bc}. The input coordinates are $(x_\text{input}=2.63\text{ cm}, y_\text{input}=-17.96\text{ cm})$. For both x and y the minimum was obtained when the proposed coordinates were close to the input coordinates, meaning that the used summary statistic preserved relevant features about the parameters of interest. (bottom left) The acquisition function resulting from the BOLFI analysis. After acquiring the set of evidence points an MCMC algorithm samples from relevant region of the parameter space. (bottom right) Contour plot and marginal posterior distributions for x and y obtained after the MCMC sampling. For both x and y an HPD $95\%$ interval is retrieved: $x \in (1.13, 4.83)$ and  $y \in (-19.56, -15.25)$.
}
\label{fig.discrepancyBOLFI}
\end{center}
\end{figure}

As pointed out in Section~\ref{bolfi}, in order to reduce the computational effort required to perform the likelihood--free analysis, the function for the discrepancies is modeled through a Gaussian process and its minimum is inferred by using Bayesian optimization. Rather than using the discrepancies directly, \cite{gutmann2016bayesian} suggested to use the logarithm of the discrepancies in cases where the underlying function is expected to be very peaked. For our study, a complication is provided by the constraint on the parameters (x,y) introduced in Equation \eqref{eq.priorABC.constraint}. We modified the acquisition function in BOLFI in order to discard those proposed coordinates for which the constraint of Equation \eqref{eq.priorABC.constraint} does not hold. A typical output provided by BOLFI is displayed in Figure \ref{fig.discrepancyBOLFI}, where the input coordinates are $(x_\text{input}=2.63\text{ cm}, y_\text{input}=-17.96\text{ cm})$. By looking at the top plot of Figure \ref{fig.discrepancyBOLFI}, it is possible to appreciate the behavior of the logarithm of the discrepancies. If the summary statistic is properly defined (i.e. is highly informative on the parameters of interest), then the minimum of the logarithm of the discrepancies will tend to the input coordinates. As a result of the Gaussian process, the target surface is inferred as shown in the bottom left plot of Figure \ref{fig.discrepancyBOLFI}. BOLFI, using any proper sampling algorithm such as MCMC or SMC, will reconstruct the posterior distribution for (x,y) by sampling from those regions of the logarithm of the discrepancies having the smallest values, as shown in the bottom plots of Figure \ref{fig.discrepancyBOLFI}.
When calling BOLFI, five of the parameters originally introduced by \cite{gutmann2016bayesian} need to be properly tuned. The first parameter, called \textit{initial evidence}, gives the number of initialization points sampled straight from the prior distributions before starting to optimize the acquisition of points. A second parameter, named \textit{update interval}, defines how often the GP hyper--parameters are optimized, while the parameter \textit{acquisition noise variance} defines the diagonal covariance of noise added to the acquired points. The fourth parameter, \textit{n evidence}, is used to build the Gaussian process to the logarithm of the discrepancies. Finally, a fifth parameter, \textit{n samples}, is used to define the length of the MCMC algorithm used to sample from the posterior distribution (the burn--in is for default equal to half of \textit{n samples}). We note that, when tuning these parameters, both inferential and computational considerations have to be taken into account. In particular, if \textit{initial evidence}, \textit{n evidence} and \textit{n sample} are fixed too small, BOLFI is faster but might fail in retrieving relevant regions of the parameter space and estimating the right variability of the posterior distribution. On the opposite, for large values of \textit{initial evidence}, \textit{n evidence} and \textit{n sample}, the computational time required by BOLFI increases.  The choices for the parameters required by BOLFI rely on the prior distributions, the employed summary statistics and the forward model used to generate the simulated sample, which make general rules for tuning the parameters required by BOLFI unfeasible. Our choices for those parameters are summarized in Table \ref{table.BOLFI.pars} and are the result of a tuning procedure that balances the inferential reliability of BOLFI (i.e. to infer relevant regions of the parameter space) with its computational time.

\begin{center}
\begin{table}[ht]
  \centering
  \begin{tabular}{*{5}{|c}|}
  \hline
  \textit{initial evidence} & \textit{update interval} &\textit{acq noise var} &  \textit{n evidence} & \textit{n sample} \\
  \hline
  \hline
  $50$ & $1$ & $5$ & $300$ & $1000$ \\
  \hline
\end{tabular}
\caption{Definition of the parameters used to initialize BOLFI. We found these choices to perform at best from both a computational and an inferential standpoint.}
\label{table.BOLFI.pars}
\end{table}
\end{center}
Once the five parameters according to Table \ref{table.BOLFI.pars} are defined, BOLFI can be used to reconstruct the  2--D (x,y) position. Considering the two different low levels for the size of the charge signal and the two different summary statistics, we perform four analyses. The location parameter from the posterior distribution used to provide the reconstructed position is the posterior mean. We tried to use also the posterior mode and the posterior median, but these two statistics did not perform as well as the posterior mean.

\begin{figure}[htbp]
\begin{center}
\includegraphics[width=1.9in]{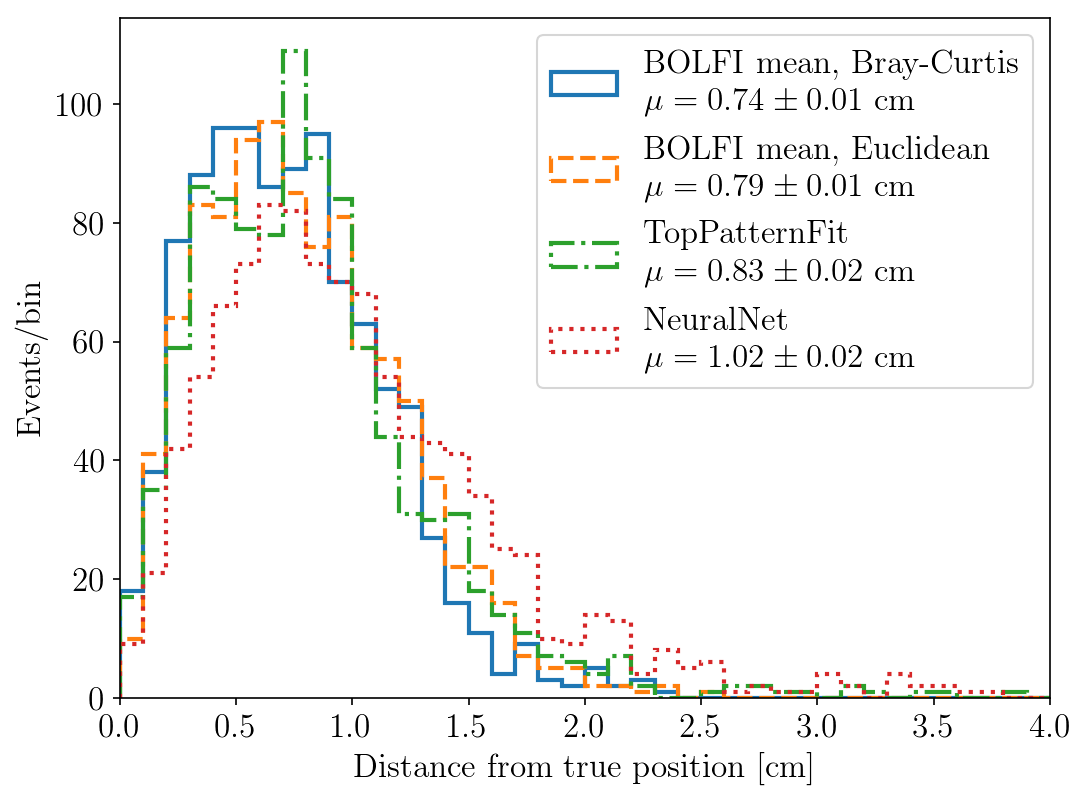}
\includegraphics[width=1.9in]{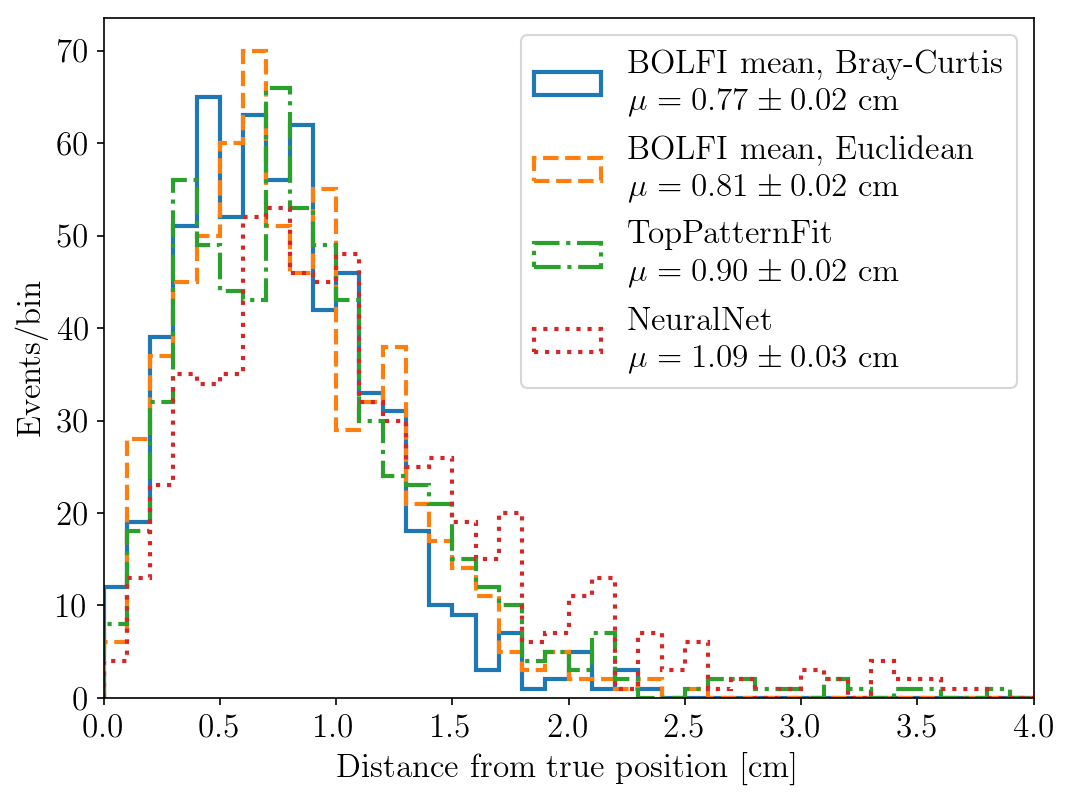}
\includegraphics[width=1.9in]{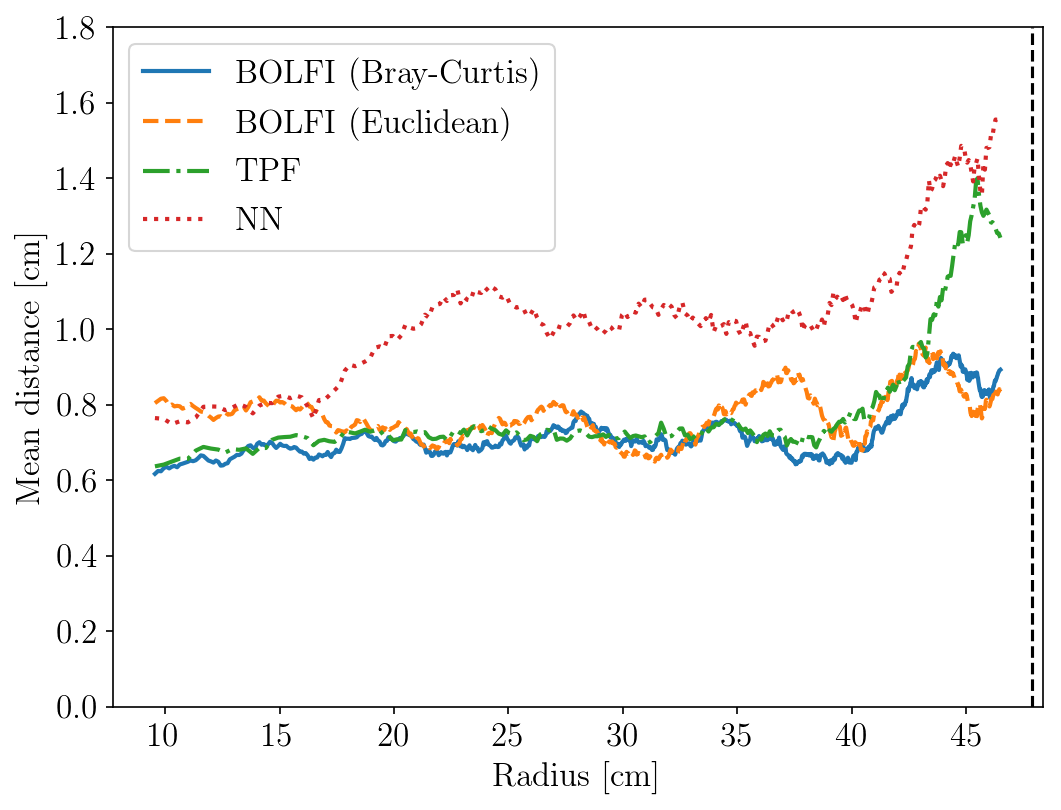}
\caption{Size of the charge signal = 25; (left) Distribution of the Euclidean distances obtained using Equation \eqref{eq.eucXY} for all the methods once $1000$ events are reconstructed. For each method the corresponding average Euclidean distance and standard deviation is displayed. When using BOLFI and the Bray--Curtis dissimilarity, the accuracy respect TPF improves of $11\%$. (middle) Distribution of the Euclidean distances obtained using Equation \eqref{eq.eucXY} for all the methods on the subset of the reconstructed events whose $R > 30\text{ cm}$. When using BOLFI and the Bray--Curtis dissimilarity, the accuracy respect TPF improves of $15\%$. The standard deviations retrieved with BOLFI are smaller than the ones obtained with the commonly employed methods. (right) Mean distances from the input positions as function of the radius $R$. When using BOLFI and the Bray--Curtis dissimilarity, the results are more accurate than TPF and NN for any level of the radius $R$.}
\label{fig:FM2.25}
\end{center}
\end{figure}

When the input size of the charge signal is fixed at $25$, BOLFI reconstructs the 2--D (x,y) position better than TPF and NN, as shown in Figure \ref{fig:FM2.25}. In particular, using the Bray--Curtis dissimilarity to compare the observed and the simulated S2 top hit pattern leads to an overall improvement, over $1000$ reconstructed positions, of $11\%$ with respect to TPF. When focusing to events on the edges of the TPC (i.e. $R > 30\text{ cm}$), the 2--D (x,y)  position reconstructed with BOLFI is $15\%$ more accurate with respect to TPF. The Euclidean distance performs slightly worse than the Bray--Curtis dissimilarity, but the results are still more accurate with respect the currently available methods ($5\%$ more accurate than TPF when focusing on all the events and $10\%$ more accurate than TPF when focusing on events at $R > 30\text{ cm}$). The right plot of Figure \ref{fig:FM2.25} shows the mean distance from the input positions as function of the radius $R$. It is possible to note that, when using BOLFI and the Bray--Curtis dissimilarity, the results are more accurate than TPF and NN for any level of the radius $R$.

\begin{figure}[htbp]
\begin{center}
\includegraphics[width=1.9in]{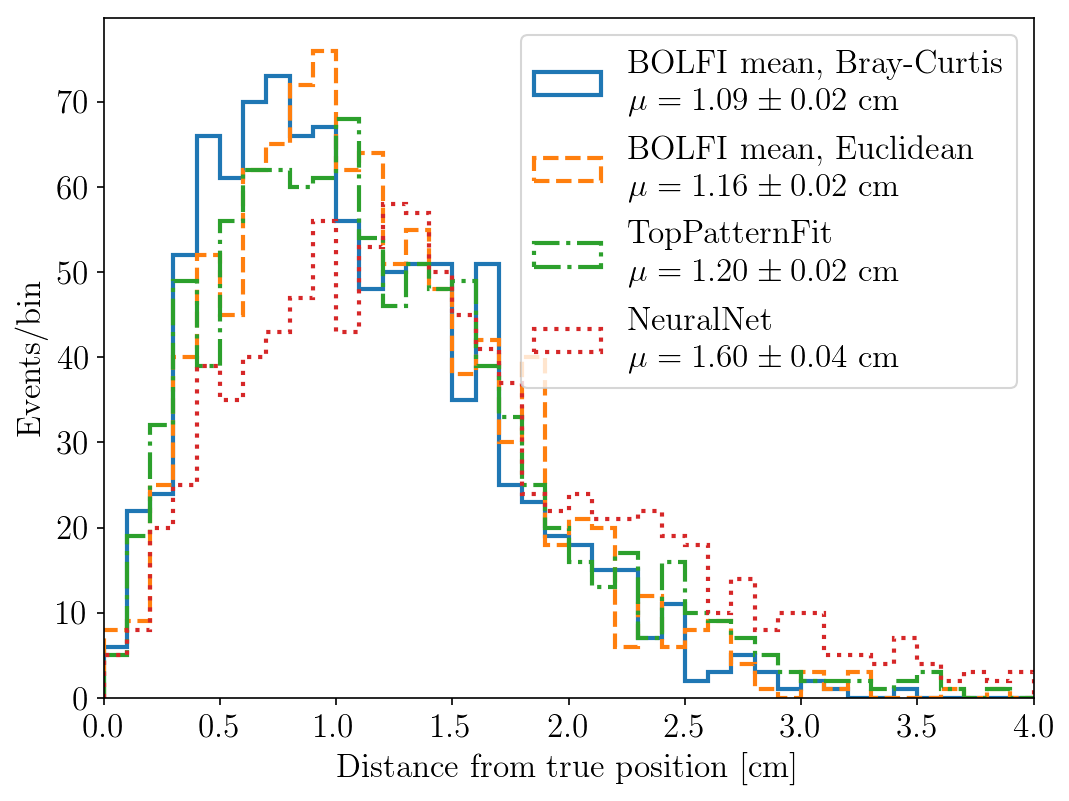}
\includegraphics[width=1.9in]{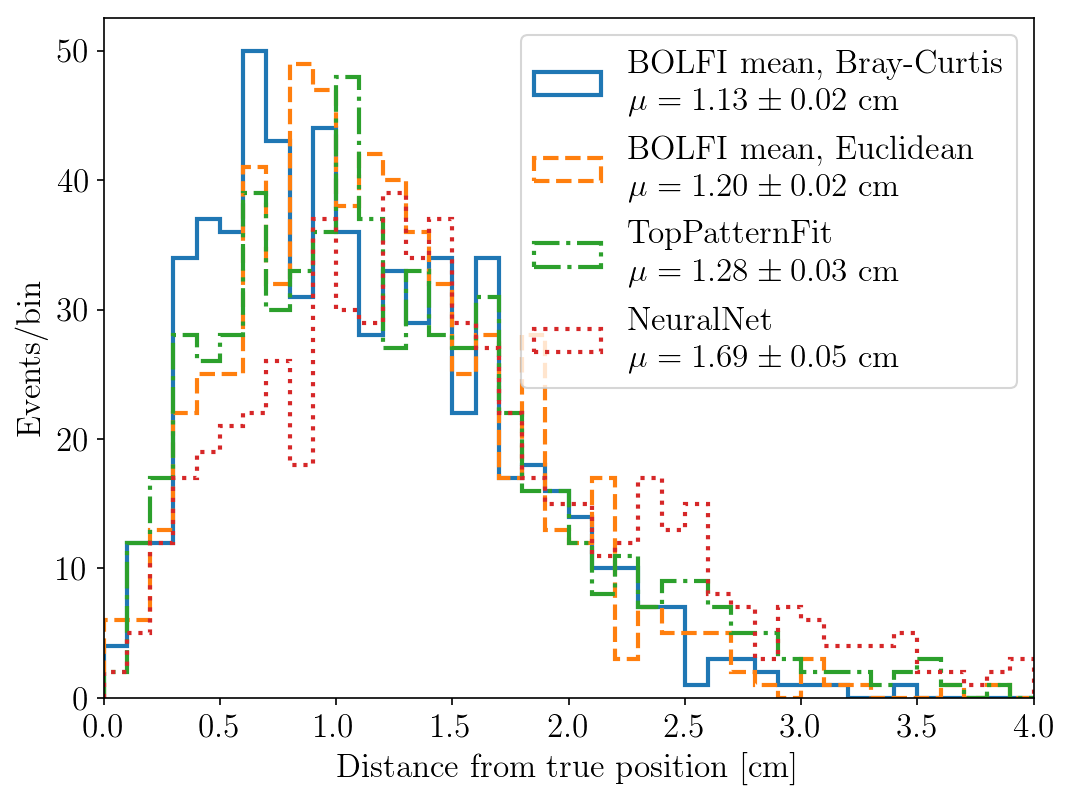}
\includegraphics[width=1.9in]{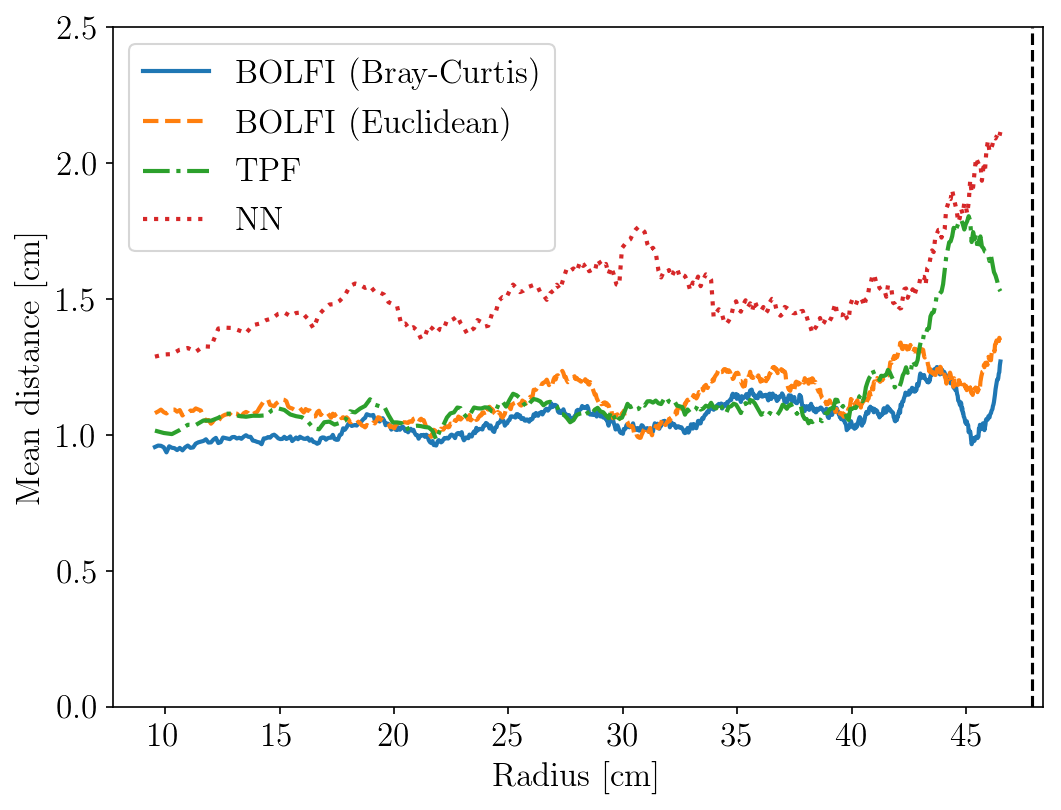}
\caption{Size of the charge signal = 10; (left) Distribution of the Euclidean distances obtained using Equation \eqref{eq.eucXY} for all the methods once $1000$ events are reconstructed. For each method the corresponding average Euclidean distance and standard deviation is displayed. When using BOLFI and the Bray--Curtis dissimilarity, the accuracy respect TPF improves of $7\%$. (middle) Distribution of the Euclidean distances obtained using Equation \eqref{eq.eucXY} for all the methods on the subset of the reconstructed events whose $R > 30\text{ cm}$. When using BOLFI and the Bray--Curtis dissimilarity, the accuracy respect TPF improves of $13\%$. The standard deviations retrieved with BOLFI are smaller than the ones obtained with the commonly employed methods. (right) Mean distances from the input positions as function of the radius $R$. When using BOLFI and the Bray--Curtis dissimilarity, the results are more accurate than TPF and NN for any level of the radius $R$.}
\label{fig:FM2.10}
\end{center}
\end{figure}

When the size of the charge signal is fixed at $10$, BOLFI also reconstructs the 2--D (x,y) position better with respect to TPF and NN, as shown in Figure \ref{fig:FM2.10}. In particular, using the Bray--Curtis dissimilarity to compare the observed and the simulated S2 hit pattern leads to an overall improvement, over $1000$ reconstructed positions, of $7\%$ with respect to TPF. When focusing to edge events at $R > 30\text{ cm}$ of the TPC, the accuracy with respect to TPF improves to $13\%$. Also in this case the Euclidean distance performs slightly worse than the Bray--Curtis dissimilarity, but the results are still more accurate than the currently available methods ($1\%$ more accurate than TPF when focusing on all the events and $7\%$ more accurate than TPF when focusing on events at $R > 30\text{ cm}$). The right plot of Figure \ref{fig:FM2.10} shows the mean distance from the input positions as function of the radius $R$. It is possible to note that, when using BOLFI and the Bray--Curtis dissimilarity, the results are more accurate than TPF and NN for any level of the radius $R$.

In order to provide statistical evidence to the qualitative evaluations discussed above, the results of the Wilcoxon Signed Rank Test, that compares the errors on the reconstruction obtained by the best BOLFI algorithm (i.e. BOLFI mean Bray--Curtis) with the errors obtained with TPF, are presented in Table \ref{table.comparisonFM2}. Defined $\Lambda = \text{median(BOLFI Bray--Curtis)} -  \text{median(TPF)}$, the null hypothesis $H_0$ stating that the medians of the errors between BOLFI mean Bray--Curtis and TPF are statistically equivalents is always rejected with P--values of order at least $10^{-7}$. Together with the P--values, we also reported the $95\%$ confidence interval for $\Lambda$. It is possible to note that all the confidence intervals do not contain $0$, suggesting that BOLFI reconstructs the ($x,y$) position statistically better than TPF, and consequently also than NN, for all the considered cases.

\begin{center}
\begin{table}[ht]
\centering
\begin{tabular}{|c|c|c|}
\hline
$\Lambda$ &  P--value ($H_0$: $\Lambda=0$)& $95\%$ confidence interval for $\Lambda$  \tabularnewline
\hline
Size of the charge signal = 25 & $1.01\cdot10^{-7}$ & $(-0.085; -0.038)$   \tabularnewline
\hline
Size of the charge signal = 25, $R > 30\text{ cm}$ &  $1.67\cdot10^{-8}$ & $(-0.12; -0.059)$   \tabularnewline
\hline
Size of the charge signal = 10 & $1.88\cdot10^{-8}$ & $(-0.11; -0.047)$  \tabularnewline
\hline
Size of the charge signal = 10, $R > 30\text{ cm}$ &  $9.25\cdot10^{-7}$ & $(-0.13; -0.056)$   \tabularnewline
\hline
\end{tabular}
\caption{Wilcoxon Signed Rank Test that compares the medians between BOLFI mean Bray--Curtis and TPF. The null hypothesis is $H_0$: the medians are statistically equivalents. The alternative hypothesis is $H_1$: TPF median is statistically smaller than BOLFI mean Bray--Curtis. For each considered case a P--value is displayed, together with a $95\%$ confidence interval.}
\label{table.comparisonFM2}
\end{table}
\end{center}

\subsection{Position and Energy Reconstruction using full waveform simulation}\label{ex.fm3}

In this last example we use similar specifications to the previously presented example but, beyond the  2--D (x,y) position, we consider also the size of the charge signal ($e$) as a parameter of interest whose posterior distribution has to be retrieved. The reconstruction of the event is defined by the 3--D coordinates ($x,y,e$). While for ($x,y$) the same prior distribution defined in the previous example is used, a prior distribution has to be assigned to the $e$. Since we are reconstructing the number of ionization electrons, the natural choice for the prior distribution is a Poisson random variable having rate parameter $\lambda$. This discrete density cannot be used in the ELFI Python package both when inferring the function of the discrepancies and when using the MCMC algorithm to retrieve samples from the posterior distributions. In order to solve this problem, we used as prior distribution for $e$ a lognormal distribution having mean the size of the charge signal reconstructed with PAX ($\text{PAX e}$) and standard deviation equal to $5$ cm:
\begin{equation}
e_\text{prop}= \text{logNormal}(\text{PAX e}, 5).
\label{eq.prior.e}
\end{equation}
Because of the presence of a third parameter, we found both the Euclidean distance and the Bray--Curtis dissimilarity defined respectively through Equations \eqref{eq.summarystat} and \eqref{eq.summarystat.bc}  to be alone not informative enough to properly reconstruct the 3--D event $(x,y,e)$. For this reason we decided to add as further distance function the energy distance, to be combined with the previously defined metrics. The energy distance is defined as:
\begin{equation}
\rho(r\obs, r\pro)_{\text{energy}} = \int_{-\infty}^{+\infty} \left(\hat{F}(r\obs) - \hat{F}(r\pro)\right)^2 d\hat{F}(r\pro),
\label{eq.summarystat.energy}
\end{equation}
where $\hat{F}(r\obs)$ and $\hat{F}(r\pro)$ are the densities estimated respectively using $r\obs$ and $r\pro$.

\begin{figure}[htbp]
\begin{center}
\includegraphics[width=5in]{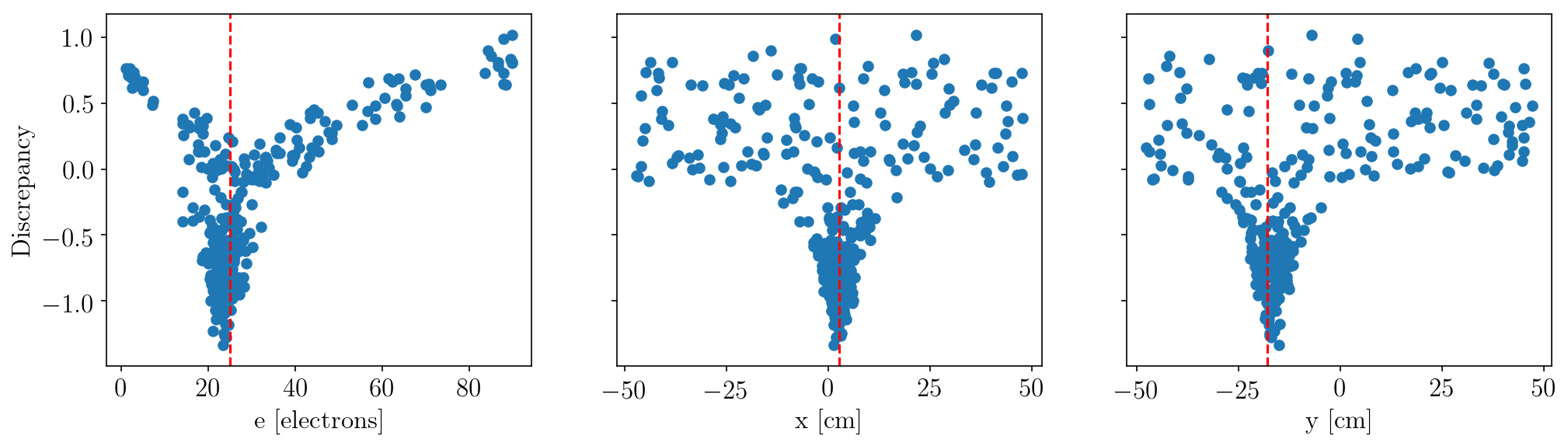}
\caption{Logarithm of the discrepancies returned by BOLFI for a generic  3--D (x,y,e) event. The input coordinates are $(e_\text{input}=25, x_\text{input}=2.63\text{ cm}, y_\text{input}=-17.96\text{ cm})$. For all three parameters the minimum was obtained when the proposed coordinates were close to the input coordinates, meaning that the used summary statistics preserved relevant features about the parameters of interest.}
\label{fig.FM3.bolfi.discrepancy}
\end{center}
\end{figure}

Although the S2 bottom hit pattern does not contain valuable information about the  2--D (x,y) position, it could be useful to better reconstruct $e$. For this reason we use as information not only the S2 top hit pattern but also the S2 bottom hit pattern. From a practical standpoint, including the S2 bottom hit pattern increases the dimension of vector returned by PAX from 127 to 248. However, we note that using the S2 bottom hit pattern to reconstruct the  2--D (x,y) position leads to worse results than using the S2 top hit pattern only, because the gain in information is smaller than the noise introduced by those further 121 PMTs. For this reason, we use both the S2 top hit pattern and  S2 bottom hit pattern to calculate the Energy Distance defined in Equation \eqref{eq.summarystat.energy} and only the S2 top hit pattern to calculate the Bray--Curtis dissimilarity defined through Equation \eqref{eq.summarystat.bc}. The overall distance function selected to compare $r\obs$ with $r\pro$ is therefore the sum between the Bray--Curtis dissimilarity that uses as information on $r\obs$ and  $r\pro$ only the S2 top hit pattern, and the Energy distance that uses as information on  $r\obs$ and  $r\pro$ all the 248 PMTs. The logarithm of the discrepancies returned by BOLFI for a generic  3--D (x,y,e) event is displayed in Figure \ref{fig.FM3.bolfi.discrepancy}. We finally note that, since in the previous example the Bray--Curtis dissimilarity performed better than the Euclidean Distance to reconstruct the  2--D (x,y) position, for this example the analyses with BOLFI only considers the Bray--Curtis dissimilarity, combined with the Energy distance as just described.

\begin{figure}[htbp]
\begin{center}
\includegraphics[width=1.9in]{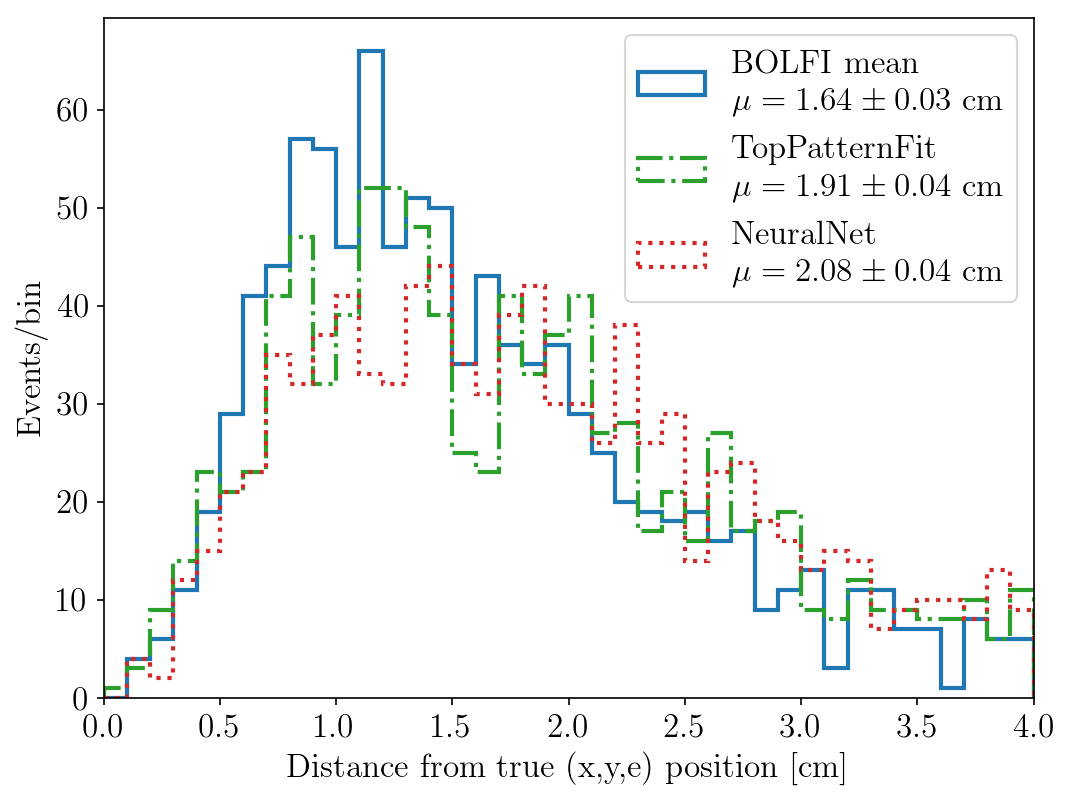}
\includegraphics[width=1.9in]{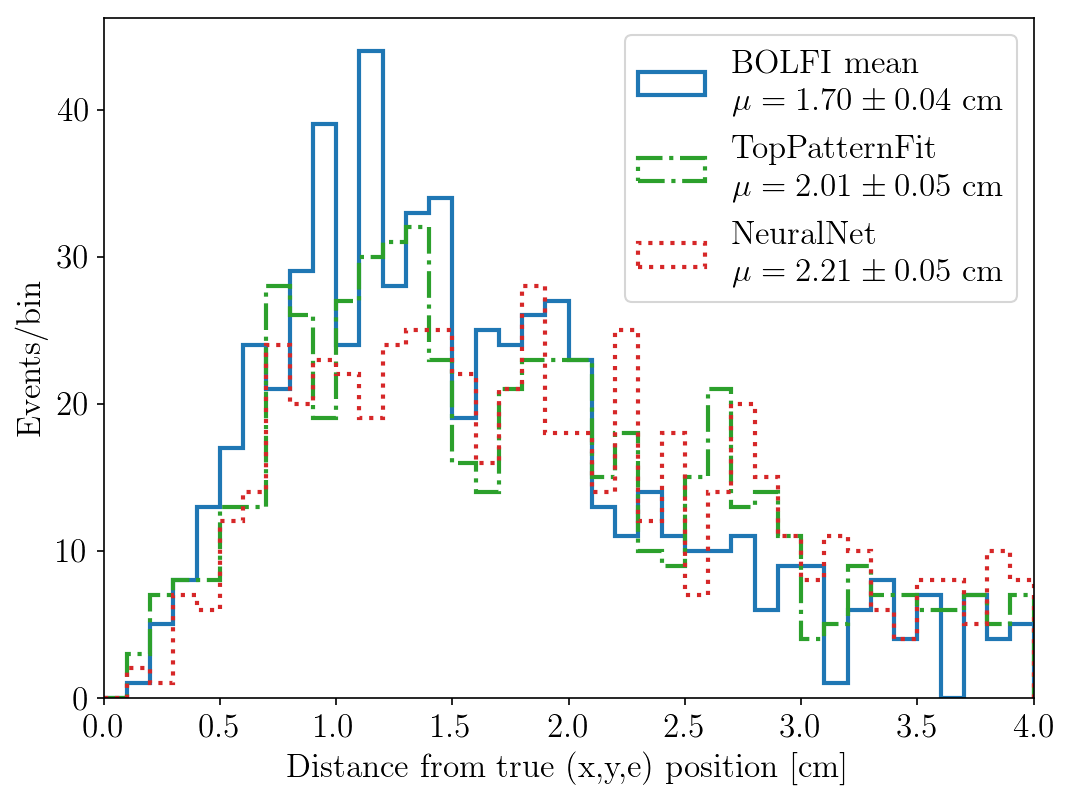}
\includegraphics[width=1.9in]{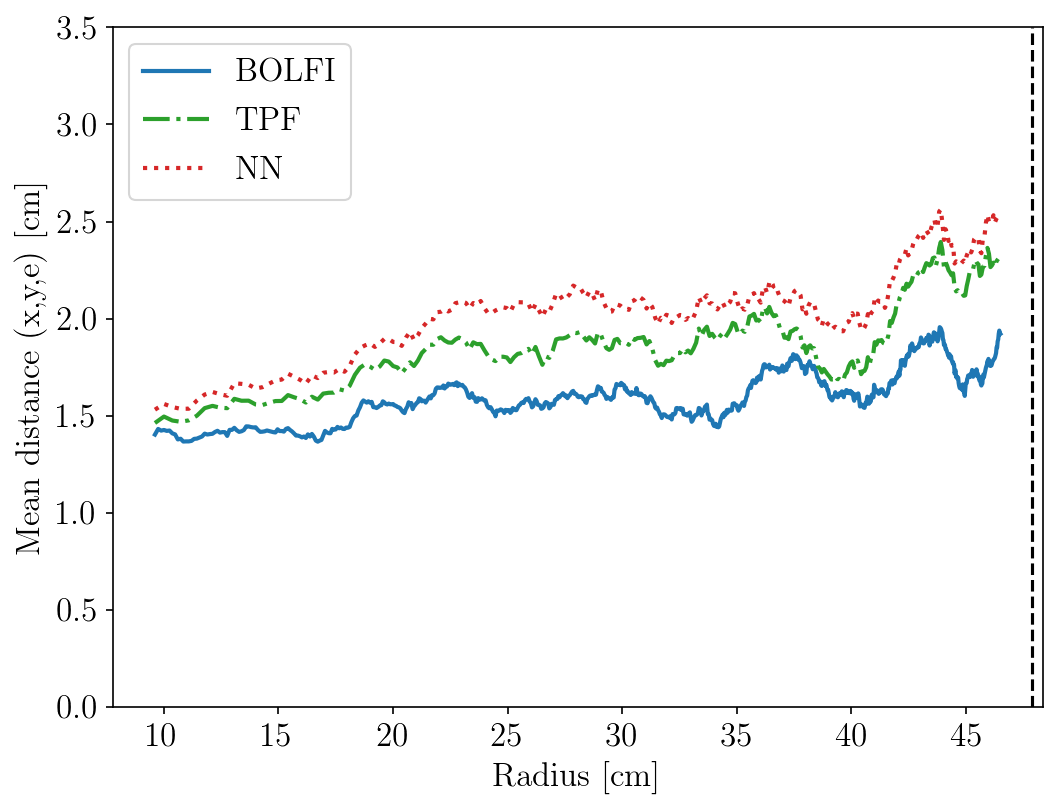}
\caption{Size of the charge signal = 25; (left) Distribution of the Euclidean distances defined in Equation \eqref{eq.eucXYE} for all the methods once $1000$ events are reconstructed. For each method the corresponding average Euclidean distance and standard deviation is displayed. When using BOLFI the accuracy respect TPF improves of $14\%$. (middle) Distribution of the Euclidean distances defined in Equation \eqref{eq.eucXYE} for all the methods on the subset of the reconstructed events whose $R > 30\text{ cm}$. When using BOLFI the accuracy respect TPF improves of $15\%$. The standard deviations retrieved with BOLFI are smaller than the ones obtained with the commonly employed methods. (right) Mean distance from the input positions as function of the radius $R$. When using BOLFI the results are more accurate than TPF and NN for any level of the radius $R$.}
\label{fig:FM3.25}
\end{center}
\end{figure}

\begin{figure}[htbp]
\begin{center}
\includegraphics[width=1.9in]{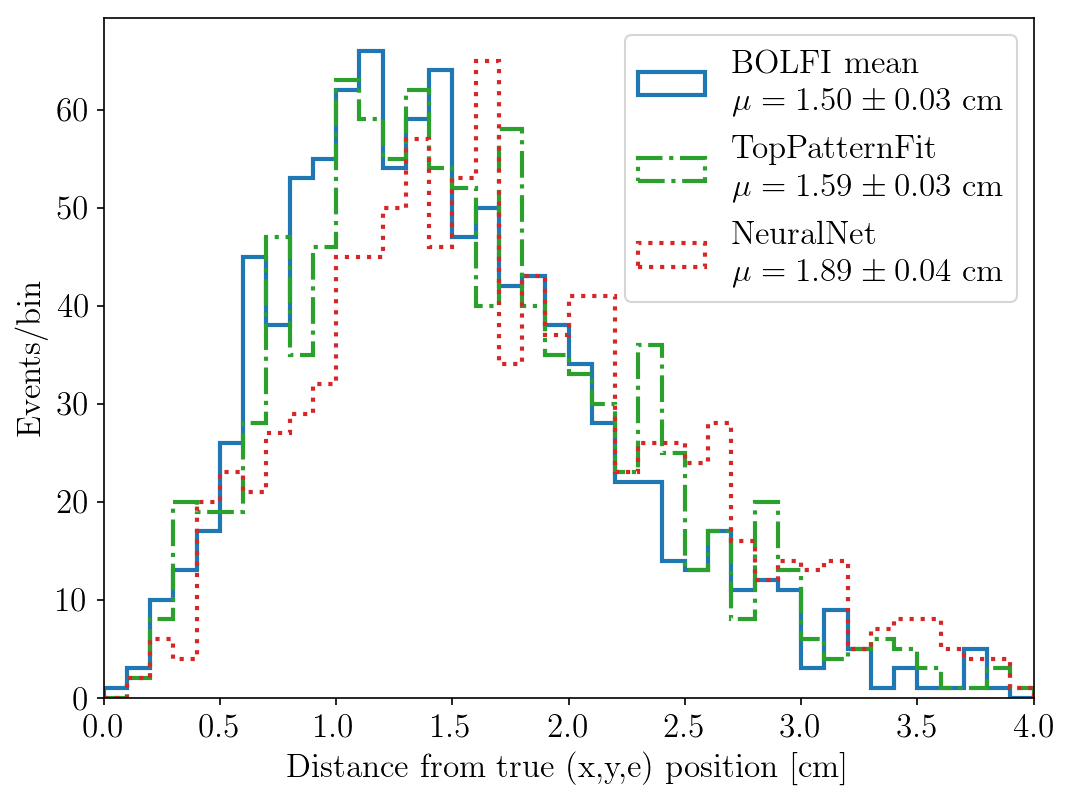}
\includegraphics[width=1.9in]{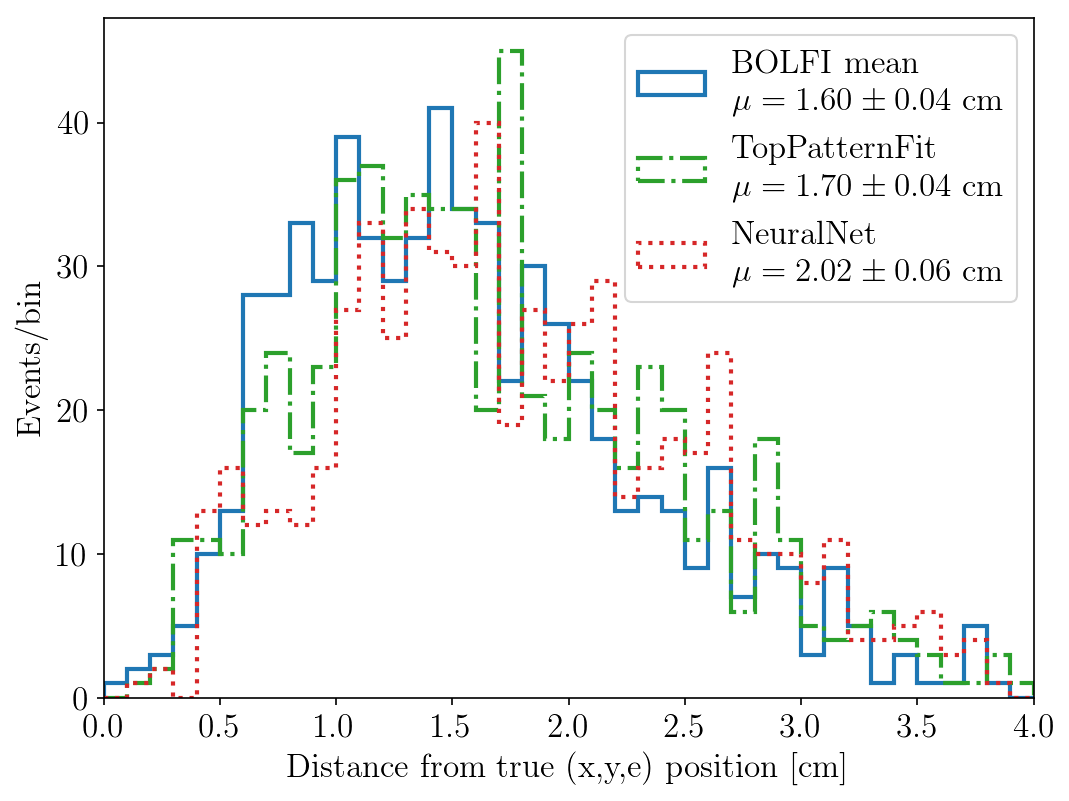}
\includegraphics[width=1.9in]{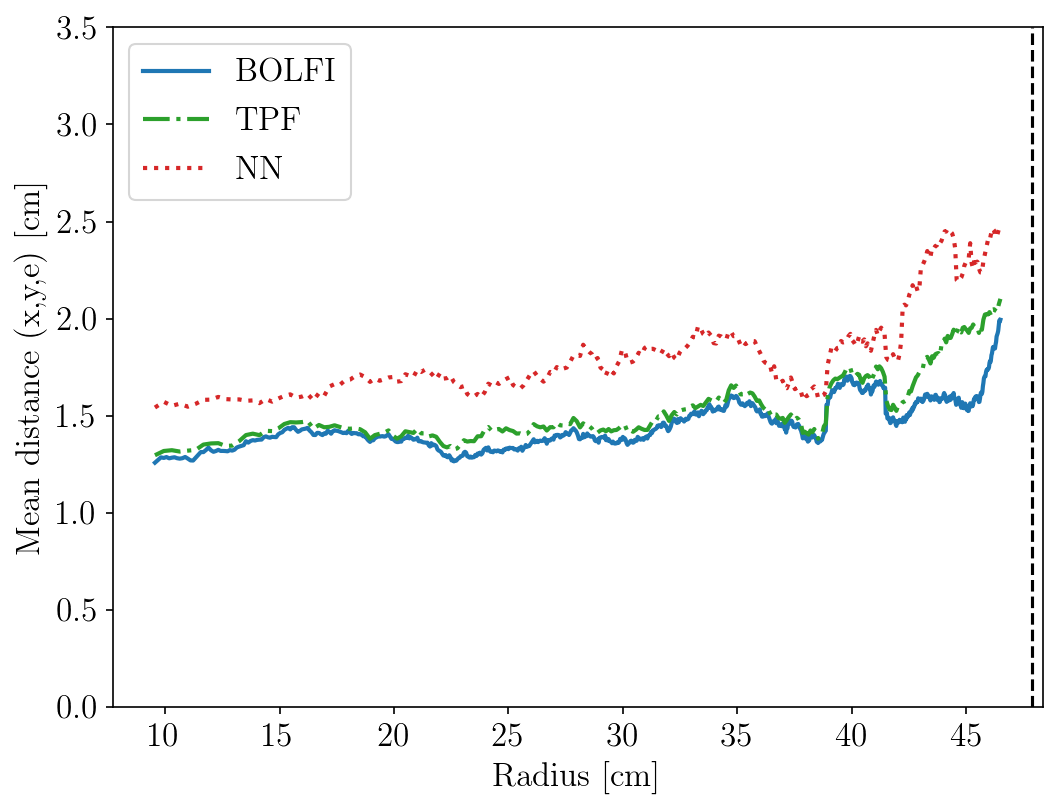}
\caption{Size of the charge signal = 10; ((left) Distribution of the Euclidean distances defined in Equation \eqref{eq.eucXYE} for all the methods once $1000$ events are reconstructed. For each method the corresponding average Euclidean distance and standard deviation is displayed. When using BOLFI the accuracy respect TPF improves of $5\%$. (middle) Distribution of the Euclidean distances defined in Equation \eqref{eq.eucXYE} for all the methods on the subset of the reconstructed events whose $R > 30\text{ cm}$. When using BOLFI the accuracy respect TPF improves of $6\%$. The standard deviations retrieved with BOLFI are smaller than the ones obtained with the commonly employed methods. (right) Mean distance from the input positions as function of the radius $R$. When using BOLFI the results are more accurate than TPF and NN for any level of the radius $R$.}
\label{fig:FM3.10}
\end{center}
\end{figure}

Figures~\ref{fig:FM3.25} and~\ref{fig:FM3.10} show the results once 1000 events have been reconstructed by using the new specifications and the settings from Table \ref{table.BOLFI.pars} required to initialize BOLFI. To quantify the goodness of the 3--D reconstruction, the Euclidean distance $d_{\text{euc}}$ in $\mathbb{R}^3$ between the reconstructed event $(x_\text{rec}, y_\text{rec}, e_\text{rec})$ and the input coordinates $(x_\text{input}, y_\text{input}, e_\text{input})$ is retrieved, according to Equation \eqref{eq.eucXYE}.
%

When the size of the charge signal is fixed at $25$ electrons, BOLFI is more accurate than TPF and NN in reconstructing the 3--D (x,y,e) coordinates, as shown in Figure \ref{fig:FM3.25}. The overall improvement, over $1000$ reconstructed positions, is of $14\%$ respect TPF. When focusing to events on the edges of the TPC (i.e. $R > 30\text{ cm}$), the accuracy respect to TPF improves of $15\%$. The right plot of Figure \ref{fig:FM3.25} shows that the results obtained with BOLFI are more accurate than TPF and NN for any level of the radius $R$.

When the size of the charge signal is fixed at $10$ electrons, BOLFI keeps being more accurate in reconstructing the 3--D (x,y,e) coordinates than TPF and NN, as shown in Figure \ref{fig:FM3.10}, although the performances are not as good as the previous case in which the size of the charge signal was equal to $25$ electrons. The overall improvement, over $1000$ reconstructed positions, is $5\%$ with respect to TPF. When focusing on events at high radii of the TPC (i.e. $R > 30\text{ cm}$), the accuracy improves of $6\%$ with respect to TPF. The right plot of Figure \ref{fig:FM3.10} shows that the results obtained with BOLFI are more accurate than TPF and NN for any level of the radius $R$.

Also for this example, in order to provide statistical evidence to the qualitative evaluations discussed above, we conclude the analyses by presenting the results of the Wilcoxon Signed Rank Test that compares the errors on the reconstruction obtained by the best BOLFI algorithm (i.e. BOLFI mean Bray--Curtis) with the errors obtained with TPF. Table \ref{table.comparisonFM3} summarizes the results of the Wilcoxon Signed Rank Test for the all the considered cases. The null hypothesis $H_0$ stating that the medians of the errors between BOLFI mean Bray--Curtis and TPF are statistically equivalents is always rejected with P--values of order at least $10^{-9}$. Together with the P--values, we also reported the $95\%$ confidence interval for $\Lambda$. It is possible to note that all the confidence intervals do not contain $0$, suggesting that BOLFI reconstruct the ($x,y,e$) events statistically better than TPF, and consequently also than NN, for all the considered cases.

\begin{center}
\begin{table}[ht]
\centering
\begin{tabular}{|c|c|c|}
\hline
$\Lambda$ &  P--value ($H_0$: $\Lambda=0$)& $95\%$ confidence interval for $\Lambda$  \tabularnewline
\hline
Size of the charge signal = 25 & $2.2\cdot10^{-16}$  & $(-0.31; -0.22)$   \tabularnewline
\hline
Size of the charge signal = 25, $R > 30\text{ cm}$ & $2.2\cdot10^{-16}$ & $(-0.34; -0.23)$   \tabularnewline
\hline
Size of the charge signal = 10 & $5.14\cdot10^{-13}$& $(-0.079; -0.045)$  \tabularnewline
\hline
Size of the charge signal = 10, $R > 30\text{ cm}$ & $4.28\cdot10^{-9}$ & $(-0.095; -0.047)$   \tabularnewline
\hline
\end{tabular}
\caption{Wilcoxon Signed Rank Test that compares the medians between BOLFI mean Bray--Curtis and TPF. Defined $\Lambda = \text{median(BOLFI Bray--Curtis)} -  \text{median(TPF)}$, the null hypothesis is $H_0$: the medians are statistically equivalents. The alternative hypothesis is $H_1$: TPF median is statistically smaller than BOLFI mean Bray--Curtis. For each considered case a P--value is displayed, together with a $95\%$ confidence interval.}
\label{table.comparisonFM3}
\end{table}
\end{center}

\section{Summary}\label{conclusions}

We presented a novel method, based on the likelihood--free framework, to reconstruct the 2--D ($x,y$) position and the 3--D ($x,y,e$) coordinates of an interaction, using the S2 signal and its corresponding S2 top and bottom hit patterns. Although for the presented examples we used the XENON1T TPC, the proposed algorithm can be used for any dual--phase TPC. In order to run both the ABC--PMC and BOLFI algorithms, we defined suitable prior distributions for the parameters of interest ($x,y,e$) and selected highly informative discrepancy measures, such as the Euclidean distance, the Bray--Curtis dissimilarity and for the last example the Energy distance.

We evaluated the quality of the proposed algorithm by reconstructing at first the 2--D ($x,y$) position and then adding as further parameter of interest the size of the charge signal $(e)$. We performed a comparison with the currently existing alternative methods on a sample of $1000$ independent events, obtained by using the open source PAX data processor and waveform simulator developed by the XENON collaboration. When focusing only on the  2--D ($x,y$) position reconstruction, BOLFI improved the accuracy of the reconstruction over TPF of $11\%$ and $7\%$, respectively when the size of the charge signal was fixed to $25$ electrons and $10$ electrons. We found even larger improvements if focusing on events having radius $R > 30 \text{ cm}$, the outer $37\%$ of the detector. For those cases BOLFI improved the accuracy of the reconstruction over TPF by $15\%$ and $13\%$, respectively when the size of the charge signal was assumed known and fixed to $25$ electrons and $10$ electrons. When focusing on the 3--D ($x,y,e$) reconstruction, BOLFI kept improving the accuracy of the reconstruction over TPF. In particular the gain in accuracy was $14\%$ and $5\%$, respectively when the input size of the charge signal was equal to $25$ electrons and $10$ electrons. Also in this more complex example we found slightly better improvements if focusing on events having radius $R > 30 \text{ cm}$, with BOLFI improving the accuracy of the reconstruction over TPF by $15\%$ and $6\%$, respectively when the input size of the charge signal was equal to $25$ electrons and $10$ electrons. Finally, the uncertainties associated to the parameters of interest retrieved by BOLFI are always the smallest among all the tested methods.

The proposed likelihood--free method presents several advantages with respect to the currently existing alternative methods. While TPF method relies on the ability to retrieve the LCE functions, that must be numerically estimated, BOLFI just requires a simulator that given a set of input coordinates provides a simulated S2 hit pattern (and as further feature the timing). BOLFI is also preferable with respect to the NN method, since it is much more accurate. We note also that BOLFI does not need any correction for reconstructing the size of the charge signal as it directly infers the number of ionization electrons. Moreover, the drift time and hence the depth coordinate $z$ can in principle be reconstructed. In the present work we used $z=0$ because simulating deep events requires a model of the drift field which is very detector specific. However if we were to reconstruct the position of deep events no field distortion correction is needed since it is included in the forward model. One possible limitation of likelihood--free inference is the computational burden of its analysis. BOLFI reconstructs events several orders of magnitude faster than the ABC--PMC algorithm but still requires order one minute reconstruction time, most of which is used by the waveform generator. The TPF method is faster, although the likelihood region is evaluated only on a $7\text{~cm }$x $7\text{~cm }$ grid. BOLFI automatically allows for properly defined credible intervals on the parameters of interest, similar to TPFs confidence intervals for the frequentist framework. Our proposed method will only be as good as the simulator can simulate the detector physics. For a real application of this method the simulator will have to be tuned to data. This was done for the PAX processor and the XENON1T TPC, but for our studies we only used simulations. The benefit of our method, compared to other reconstruction algorithms, is what will allow for a complete simulator which can simulate many physical effects. For example the PAX processor does simulate field distortion from the non--uniformities in the electric drift field in the TPC which was tuned to data. Ultimately the speed and accuracy of the simulator determines the speed and accuracy of the reconstruction. We note however that the gain in accuracy presented in this work justifies the applicability of BOLFI over TPF even if our algorithm is computationally more expensive.

\acknowledgments
We gratefully acknowledge and thank the XENON collaboration for developing and maintaining their open source data processor and waveform generator PAX. U. Simola was supported by the Academy of Finland grant no. 1313197. J. Conrad acknowledges support from the Knut and Alice Wallenberg Foundation and Swedish Research Council. J. Corander was supported by the ERC grant no. 742158.

\bibliographystyle{unsrt}
\bibliography{biblio}

\end{document}